\documentclass[manuscript,screen]{acmart}

\usepackage{multirow}    					
\usepackage{hhline}     				 		
\usepackage{algorithm} 
\usepackage{algpseudocode}
\usepackage{enumerate}
\usepackage{wrapfig}
\usepackage{fancybox}
\usepackage{url}
\usepackage{balance}

\usepackage{amssymb,amsmath}
\usepackage{pifont}
\usepackage{xcolor}
\usepackage{tikz}
\usepackage{graphicx}
\usepackage{color}    
\usepackage{listings}
\usepackage{siunitx}
\usepackage{soul} 
\usepackage[inline]{enumitem} 

\definecolor{KWColor}{rgb}{0.37,0.08,0.25}
\definecolor{CommentColor}{rgb}{0.133,0.545,0.133}
\definecolor{StringColor}{rgb}{0,0.126,0.941}
\usepackage{subfigure}
\usepackage{algorithm} 
\usepackage{algpseudocode}
\usepackage{listings}
\usepackage{upquote}
\usepackage{float}

\usepackage{amssymb,amsmath}
\usepackage{ifthen}
\usepackage{xcolor}
\usepackage{soul}
\soulregister{\cite}7
\soulregister{\citep}7
\soulregister{\citet}7
\soulregister{\ref}7 
\soulregister{\pageref}7 
\usepackage[page,toc,titletoc,title]{appendix}
\usepackage{longtable}

\lstdefinestyle{JAVA}{
  language=JAVA,
  moredelim=[is][\underbar]{_}{_},
}

\usepackage[most]{tcolorbox}

\newcommand\tool[1]{HiSenDroid}

\setcopyright{acmcopyright}
\copyrightyear{2022}
\acmYear{2022} \acmMonth{10}

\begin{document}

\title{Demystifying Hidden Sensitive Operations in Android apps}

\author{Xiaoyu Sun}
\email{xiaoyu.sun@monash.edu}
\affiliation{
  \institution{Monash University, Australia}
  \streetaddress{Wellington Rd}
  \city{Clayton}
  \state{VIC}
  \postcode{3800}
}

\author{Xiao Chen}
\email{Xiao.chen@monash.edu}
\affiliation{
  \institution{Monash University, Australia}
  \streetaddress{Wellington Rd}
  \city{Clayton}
  \state{VIC}
  \postcode{3800}
}

\author{Li Li}
\authornote{Li Li is the corresponding author.}
\email{lilicoding@ieee.org}
\affiliation{
  \institution{Monash University, Australia}
  \streetaddress{Wellington Rd}
  \city{Clayton}
  \state{VIC}
  \postcode{3800}
}

\author{Haipeng Cai}
\email{haipeng.cai@wsu.edu}
\affiliation{
  \institution{Washington State University, United States}
  \streetaddress{Wilson Rd}
  \city{Pullman}
  \state{WA}
  \postcode{99164-5910}
}

\author{John Grundy}
\email{john.grundy@monash.edu}
\affiliation{
  \institution{Monash University, Australia}
  \streetaddress{Wellington Rd}
  \city{Clayton}
  \state{VIC}
  \postcode{3800}
}

\author{Jordan Samhi}
\email{jordan.samhi@uni.lu}
\affiliation{
  \institution{University of Luxembourg, Luxembourg}
  \streetaddress{2 Avenue de l'Universite, 4365 Esch-sur-Alzette, Luxembourg}
}

\author{Tegawend\'e F. Bissyand\'e}
\email{tegawende.bissyande@uni.lu}
\affiliation{
  \institution{University of Luxembourg, Luxembourg}
  \streetaddress{2 Avenue de l'Universite, 4365 Esch-sur-Alzette, Luxembourg}
}

\author{Jacques Klein}
\email{jacques.klein@uni.lu}
\affiliation{
  \institution{University of Luxembourg, Luxembourg}
  \streetaddress{2 Avenue de l'Universite, 4365 Esch-sur-Alzette, Luxembourg}
}

\begin{abstract}
Security of Android devices is now paramount, given their wide adoption among consumers. 
As researchers develop tools for statically or dynamically detecting suspicious apps, malware writers regularly update their attack mechanisms to hide malicious behavior implementation. 
This poses two problems to current research techniques: static analysis approaches, given their over-approximations, can report an overwhelming number of false alarms, while dynamic approaches will miss those behaviors that are hidden through evasion techniques. 
We propose in this work a static approach specifically targeted at highlighting hidden sensitive operations, mainly sensitive data flows. The prototype version of HiSenDroid has been evaluated on a large-scale dataset of thousands of malware and goodware samples on which it successfully revealed anti-analysis code snippets aiming at evading detection by dynamic analysis. We further experimentally show that, with FlowDroid, some of the hidden sensitive behaviors would eventually lead to private data leaks.
Those leaks would have been hard to spot either manually among the large number of false positives reported by the state of the art static analyzers, or by dynamic tools. Overall, by putting the light on hidden sensitive operations, HiSenDroid helps security analysts in validating potential sensitive data operations, which would be previously unnoticed.

\end{abstract}



\maketitle
\section{Introduction}
Android is the most adopted mobile operating systems in terms of users, applications and developers~\cite{IDCReport}. However, its popularity means that legitimate developers must co-exist with malware writers. Reports on many different kinds of attacks are presented in the technology and lay media. For example, security researchers have reported a malicious ``clicker trojan''\footnote{Such as the \emph{Android.Click.312.origin} trojan and its modified variant {Android.Click.313.origin} trojan. This aims to generate fraudulent click-through and subscription revenues.} which has been bundled into 34 different Google Play apps that have already been installed more than 100 million times\footnote{\url{https://www.forbes.com/sites/zakdoffman/2019/08/13/android-warning-100m-users-have-installed-dangerous-new-malware-from-google-play/\#1956f51c22a9}}. On a larger scale, antivirus engines have been flagging a large number of apps as potential threats. For example, as of October 2020, the popular AndroZoo dataset~\cite{allix2016androzoo} has recorded more than 226,000
Android GooglePlay apps than have been flagged as adware/malware by at least 5 Antivirus products, and this number is still growing.
Those adware/malware often not work along but collaborate with many third parties over the internet. Some of the representative malicious behaviors include leading users to malicious websites through devious advertisements~\cite{liu2020maddroid, 9282795, dong2018frauddroid, dong2018mobile}, distributing malicious apps in the mobile network through drive-by downloads~\cite{cova2010detection}, leaking users' sensitive data to web servers through HTTP connections~\cite{sun2021characterizing,liu2021first,li2015potential,gao2020borrowing}, etc.

To protect Android users against the rapid spread of malware, the research and practice communities have implemented a variety of measures and proposed several approaches to detect malware~\cite{arp2014drebin, mariconti2016mamadroid, zhao2021impact,li2017understanding,sun2022mining,xu2022lie}. These include static code analysis-based approaches~\cite{li2017static,li2019rebooting}, dynamic testing based approaches~\cite{kong2018automated}, and learning-based approaches~\cite{liu2022deep}. 
Unfortunately the emergence of many different malware detection techniques has also stimulated malware attackers into being more innovative to increasingly better hide malicious behaviour, in order to bypass static code analysis (e.g., via obfuscation) and even dynamic detection (e.g., sensing of sandbox execution). 
In practice, sophisticated code obfuscation techniques \cite{moser2007limits} are being leveraged by attackers to hide their malicious program behavior, leading to false negatives in most static analyses thus resulting in imprecise and unsound results.
Camouflage techniques have been frequently leveraged by attackers to evade dynamic testing approaches~\cite{rasthofer2017making,egele2008survey}.
Attackers often introduce a so-called logic bomb or time bomb to set off malicious functions only after certain conditions are met.
For instance, after knowing that Google  applies a dynamic analysis tool called \emph{bouncer} to scan every app submitted to Google Play for five minutes, as revealed by Oberheide et al.~\cite{oberheide2012dissecting}, a bunch of malicious apps has been created and been demonstrated to be capable of penetrating Google's bouncer vetting system by simply waiting five minutes before triggering their malicious behavior. 

To cope with such hidden malicious behaviors, researchers have devised new detection approaches.
For example, Fratantonio et al.~\cite{fratantonio2016triggerscope} have proposed an approach called TriggerScope to detect hidden behaviors triggered by predefined circumstances such as events related to location, time, and SMS. 
However, TriggerScope is not capable of detecting such malicious activities hidden behind other trigger types, such as the existence of other services (i.e., other than location, time and SMS).
In line with this research, Pan et al.~\cite{pan2017dark} have proposed a machine learning-based approach aiming to discover unknown trigger types.
Their approach, however, needs to manually label a dataset for training, which is known to be resource-intensive and error-prone.

Static analyzers suffer less than dynamic approaches from evasion techniques such as logic bomb or time bomb. 
In particular, regarding sensitive flow detection (also called privacy leak detection), numerous static analysis tools have been proposed such as FlowDroid~\cite{arzt2014flowdroid} (and its extension \textsc{IccTA}~\cite{li2015iccta}), \textsc{Amandroid}~\cite{wei2014amandroid},  or \textsc{DroidSafe}~\cite{gordon2015information}. Although these tools are able to track sensitive flows (which are often hidden) by bringing  key new contributions to the research community, they still face some well-known limitations~\cite{AnalyzingAnalyzers-ISSTA2018}: their inherent over-approximations inevitably lead to false alarms, which, for some analyzers, occur at a high rate, making them impractical. Consequently, when building on static analysis, manual investigation is often required.
Unfortunately, such efforts cannot scale. Dynamic validation then appears as an alternative. Unfortunately, runtime execution often misses hidden sensitive flows due to the implementation of evasion techniques by attackers.
While some effort (e.g.,~\cite{fratantonio2016triggerscope,pan2017dark}) has been put to \emph{characterize} Hidden Sensitive Operations (HSOs) in Android apps, our community has not yet proposed dedicated approaches to \emph{detect and explain} such operations, allowing attackers to achieve malicious behaviors while bypassing certain security vetting mechanisms.

We fill this research gap in this work by proposing a new prototype tool, \tool{}, which deploys an automated static app analyzer tailored for detecting \textit{hidden} sensitive operations. 
\tool{} performs a sequence of static analyses, including call graph analysis, forward data-flow analysis, inter-procedural backward data-flow analysis, etc. 
For exposed HSOs, \tool{} further goes one step deeper to record detailed information for explaining why these HSOs should be flagged as such.

To summarize, key contributions of our work include:
\begin{itemize}
    \item
    We propose using a static analysis approach to discover hidden sensitive operations that are not exposed to the state-of-the-art static and dynamic analysis tools in Android apps. To this end, we leverage control flow and data flow analyses to identify the unique code level characteristics of hidden sensitive operations.
    \item
    We designed and implemented a prototype tool \tool{} for analyzing hidden sensitive operations. We release \tool{} as an open source project \cite{HiSenDroid} for supporting security analysts in their analysis needs and fostering further researches in this direction.
    \item We evaluated \tool{} on a large-scale dataset that contains 10,000 benign and 10,000 malware samples, and discovered emerging anti-analysis techniques employed by malware samples, such as fulfilling certain restrictions related to \emph{time}, \emph{location}, \emph{SMS message}, \emph{system properties}, \emph{package manager}, and other logics. 
    \item With the help of FlowDroid~\cite{arzt2014flowdroid}, a static taint analyzer, we further experimentally show that HSOs have been recurrently leveraged by attackers to leak sensitive user information. 
\end{itemize}

The rest of the paper is organized as follows: Section \ref{sec:motivation} defines HSO and presents the motivation of our research, i.e., why there is a strong need to demystify HSO.
Section~\ref{sec:approach} depicts the design and implementation of the proposed approach. Section~\ref{sec:commonHSO} and Section~\ref{sec:evaluation} respectively describe the characteristics of common and susipious HSOs detected by our approach from a large-scale dataset.
Section~\ref{sec:implication} presents a practical implication of our approach by characterizing sensitive data leaks triggered by HSOs.
Section \ref{sec:limitations} discusses the limitations of the tool. 
Section \ref{sec:related_work} reviews the related works, and finally Section \ref{sec:conclusion} concludes this paper.

\section{HSO Definition and Motivation}
\label{sec:motivation}

We conducted an exploratory study to understand the characteristics of \emph{Hidden Sensitive Operations} (HSO) in Android apps. We first dumped operations in a set of real-world Android malware. Then, we manually examined those operations to observe the characteristics of such operations that could be considered as hidden-triggered operations.
Based on our manual summarization, we found that
(1) \emph{if statement} and the notion of \emph{branch} are key in the definition of HSO; 
(2) the \emph{if statement} contains a specific \emph{operation} that triggers the hidden sensitive flows, and this trigger condition is related to Android API.

Let $B$ denote one of the two branches of an \emph{if-then-else statement}, or the branch of an \emph{if statement} where the \emph{else} branch is considered empty.

\textbf{Definition 1 [Hidden Sensitive Branch (HSB)]:} 
$B$ is an HSB if it fulfills the following \textbf{rules}: 
\begin{enumerate}
    \item $B$ contains sensitive Android APIs, and these APIs are different from those contained in the other branch involved in the \emph{if-then-else statement}. 
    The rationale behind this condition is that a hidden branch is supposed to achieve some sensitive behaviors that are different from those of the "normal" branch (i.e., non-HSB), which per se might also access sensitive APIs as part of the app's expected behaviors.
    \item $B$ does not involve any of the variables appearing in the \emph{condition expression} of the \emph{if-then-else statement}. The rationale behind this is that the branch is triggered by conditions that are also different from its (sensitive) behaviors.
\end{enumerate}

Less formally, an HSB could be defined as an "if branch" which accesses sensitive APIs, and which is fully "independent" of the \emph{if condition} and the other branch of the  \emph{if statement}. 

Let $C$ denote the $condition$ of an \emph{if statement}.

\textbf{Definition 2 [Hidden Sensitive Operation (HSO)]:} 
An HSO is an HSB that is triggered by a condition $C$ containing values obtained via (or directly impacted by) Android system APIs or system properties (i.e., attributes of system classes). This may return different values when being executed under different circumstances, so as to triggering hidden sensitive operations.

\begin{lstlisting}[
caption={An example of a real-world hidden sensitive data flow.},
label=code:emu,
float=t,
firstnumber=1]
public class MainActivity extends AppCompatActivity {
 protected void onCreate(Bundle savedInstanceState) {
  SmsManager smsManager = SmsManager.getDefault();
  ED ed = new ED(this);
  StringBuilder message = new StringBuilder();

  if(ed.checkPackageName()) {
   TelephonyManager tm = (TelephonyManager)     getSystemService(Context.TELEPHONY_SERVICE);
   String imei = tm.getDeviceId();
   String phoneNumber = tm.getLine1Number();
   String subscriberId = tm.getSubscriberId();
   message.append(imei);
   message.append(phoneNumber);
   message.append(subscriberId);
   smsManager.sendDataMessage("+115800763861", null, (short)1001, message.toString().getBytes(), null, null);
  } else {
   //benign string operations
}}


public class ED {
 public ED(Context pContext) {
  mContext = pContext;
  mListPackageName.add("com.google.android...genymotion");
  mListPackageName.add("com.bluestacks");
  mListPackageName.add("com.bignox.app");
 }
 public boolean checkPackageName() {
  if (!isCheckPackage || mListPackageName.isEmpty()) {
   return false;
  }
  final PackageManager packageManager = mContext.getPackageManager();
  for (final String pkgName : mListPackageName) {
   final Intent tryIntent = packageManager.getLaunchIntentForPackage(pkgName);
   if (tryIntent != null) {
    final List<ResolveInfo> resolveInfos = packageManager.queryIntentActivities(tryIntent, PackageManager.MATCH_DEFAULT_ONLY);
    if (!resolveInfos.isEmpty()) {
     return true;
    }
   }
  }
  return false;
}
\end{lstlisting}

Listing 1 exemplifies a simplified code snippet illustrating these definitions in practice. Note that Listing 1 presents the typical characteristics of an HSO in many real-world apps that we have manually analyzed. 
At line 7, the app firstly checks if it is running on one of the popular Android emulators (i.e., \emph{genymotion, bluestacks, and bignox}). If not, the app reads the device information and sends it to a hard-coded phone number through an SMS. Otherwise, if an emulator environment is detected, it will only perform some unharmful string operations (ignored).
In this example, three private data -- namely the device's IMEI, IMSI, and phone number -- are retrieved in lines 9-11 and sent to a hard-coded phone number via SMS (line 15).
All of these three leaks are hidden behind the trigger condition \emph{ed.checkPackageName()} (line 7). The trigger condition checks the return value of a self-defined method \emph{checkPackageName()} (line 30), which is determined by several other \emph{if-conditions} defined in the invoking method (lines 31,37,39). Finally, the trigger condition in the HSO is traced back to a system API \emph{PackageManager.queryIntentActivities()} (line 38) (\textbf{cf. Definition 2}).
This trigger condition examines whether popular Android emulator packages (lines 26-28) are available in the device, i.e., checking if the app is running on these emulators. If the running environment is not one of the hard-coded emulators, the HSO will be performed.
Otherwise, benign string operations are executed (lines 17-19) (\textbf{cf. Definition 1}).

\section{Our Approach}
\label{sec:approach}

\begin{figure}[!h]
    \centering
    \includegraphics[width=0.8\linewidth]{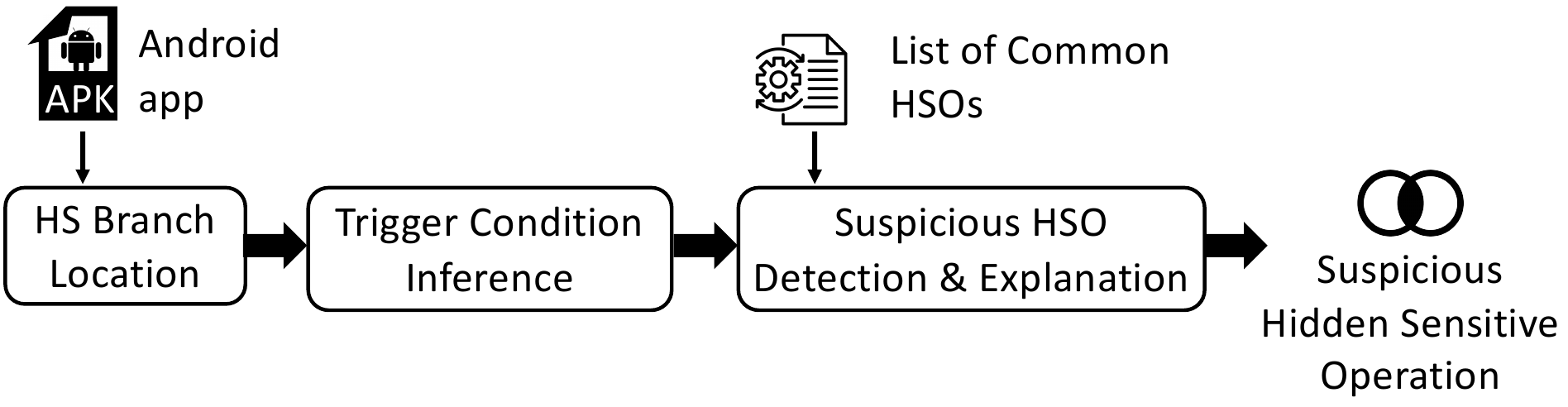}
	\caption{The working process of \tool{}.}
    \label{fig:overview}
\end{figure}

To better help security analysts understand Hidden Sensitive Operations (HSO) placed in Android apps, we designed and implemented a prototype tool, named \tool{}, to automatically locate such operations in Android apps.
\tool{} takes as input an Android app and outputs a set of hidden sensitive operations.
Fig.~\ref{fig:overview} illustrates the working process of \tool{}. It achieves the aforementioned goal through three main modules, namely: (1) Hidden Sensitive Branch Location; (2) Trigger Condition Inference; (3) Suspicious HSO Detection and Explanation. 
We now respectively detail these three modules.

\subsection{Hidden Sensitive Branch Location}
\label{subsec:hbl}

The first module of \tool{} is responsible for locating hidden sensitive branches (HSBs) in Android apps (i.e., fulfilling the rules in Definition 1).
Towards locating HSBs, this module first statically goes through all the methods that appeared in the DEX file of the input APK.
For each method, this module then constructs an intra-procedural control-flow graph (CFG) and traverses the graph to locate \emph{if-then-else statements}.
Once an \emph{if-then-else statement} is located, it further extracts the sensitive APIs accessed by the two branches 
(hereinafter referred to as \emph{if-branch} and \emph{else-branch}).
Sensitive APIs are such methods that are protected by Android permissions, which are classified following the latest Android API-permission mappings PSCout~\cite{au2012pscout}, Axplorer~\cite{backes2016demystifying}, Arcade~\cite{aafer2018precise}, and NatiDroid~\cite{chaoran2022cross}.
Any of the two branches will be considered a 
potential HSO if it has indeed accessed sensitive APIs that
are different from the APIs accessed by the other branch. 

When extracting sensitive APIs, in order to obtain a \emph{soundy} result~\cite{livshits2015defense} (e.g., including all the sensitive APIs accessed by a potential HSB), this module traverses not only the methods directly presented in the HSB but also all the methods that could be reached from the branch.
This process is made possible by first constructing a call graph (CG) for the input APK.
Unfortunately, as discussed by many existing works, Android apps do not have a single entry point (e.g., \emph{main()}) that connects other parts of the application code, making static analyses challenging to cover all the app code. 
Fortunately, this challenge has been well addressed by the state-of-the-art by artificially creating a so-called dummy main method, connecting together all the separated code parts, including system-driven lifecycle methods and event-driven callback methods \cite{arzt2014flowdroid}. 

Based on our observation and the findings of previous work~\cite{pan2017dark}, the connection between trigger conditions and the operations along its paths is often weak. Indeed, the variables appearing in triggers typically do not propagate data flow to its following paths. Take Listing 1 as an example, the app checks if it is running inside Android emulators at line 7, where the trigger condition code itself is not supposed to steal private data and is only meant to determine the right situation for running hidden code. To leverage this property, we attempt to check whether variables appearing in the HSB have data dependency with any variable within the condition expression.
Thus, for a given potential HSB, this module goes one step further to check if any of the variables appeared in the HSB's \emph{condition expression} has been leveraged by the HSB code.
If so, this HSB will not be considered as a true HSB and thereby will be excluded from further analyses.
This module achieves this by conducting a simple intra-procedural control-flow analysis.
In a case of true HSB, there should not be intersections between the set of variables that appeared in its \emph{condition expression} and those within the branch.

\subsection{Trigger Condition Inference}

After locating HSBs, the second module goes one step deeper to infer hidden sensitive operations (HSOs) so as to fulfill Definition 2.
Given a true HSB, the idea of detecting HSOs is to infer the detailed trigger conditions that lead to the execution of the HSB. 

We began with a preliminary study to understand what kinds of trigger conditions have been used to hide suspicious APIs, as identified from the literature~\cite{pan2017dark, petsas2014rage, chen2008towards, vidas2014evading, jing2014morpheus, diao2016evading, costamagna2018identifying, Android.hehe, Hackingteam, tamperingdetection, norboev2017robustness} on trigger conditions. 
For example, Petsas et al.~{\cite{petsas2014rage}} investigated anti-analysis techniques that can be employed by Android apps to evade detection, including pre-initialized static information(e.g., IMEI value), dynamic information that does not change (e.g., Sensors data) and VM instruction emulation (e.g., hardware variable). 
In their paper, they demonstrated how dynamic analysis could be evaded by the aforementioned trigger conditions in an emulated environment. 
Pan et al.~{\cite{pan2017dark}} further summarize that almost all the trigger conditions of HSOs can be characterized by \textbf{System Properties} (e.g., OS or hardware traces of a mobile device) or \textbf{Environment Parameters} (time, locations, SMS, etc.).
To the best of our knowledge, the values in both types can be obtained through Android system APIs.
In other words, an HSO trigger condition is expected to involve, directly or indirectly,
one or more system API calls for interacting with the Android operation system.
Therefore, since it is very important to identify all possible trigger conditions, we propose considering all the condition checks to infer HSO's trigger conditions\footnote{We remind the readers that state-of-the-art studies (e.g., by Moser et al.~\cite{moser2007limits} and Zeng et al.~\cite{zeng2018resilient}) have further revealed that obfuscation (via reflective calls or opaque predicates) could be leveraged to complicate the inference of trigger conditions (e.g., changing the way how a system property is obtained from the system). We do not take obfuscation as a type of trigger condition but will only consider it as a technique that complicates the process of identifying trigger conditions. We will discuss the impact of obfuscation on our approach at the end of Section~\ref{sec:evaluation}.} as long as they involve system properties, environment parameters, and any other values yielded by system APIs.

In this work, we follow the same criteria to infer HSOs (i.e., the trigger conditions involves values obtained through Android system APIs).
Specifically, to infer the trigger conditions, for each of the variables that appeared in the HSB's \emph{condition expression}, there is a need to conduct backward data-flow analyses to locate its definition statement.
The code block between the definition statement and the \emph{if-then-else statement} is then referred to as a Condition Triggering Block (CTB).
Then, given a potential HSO, we check whether a system API is involved in the \emph{definition statement} of the HSO's CTB.
If not, we will regard this HSO as a false result and consequently will not consider it for further analyses.

\begin{figure}[!h]
    \centering
    \includegraphics[width=0.6\textwidth]{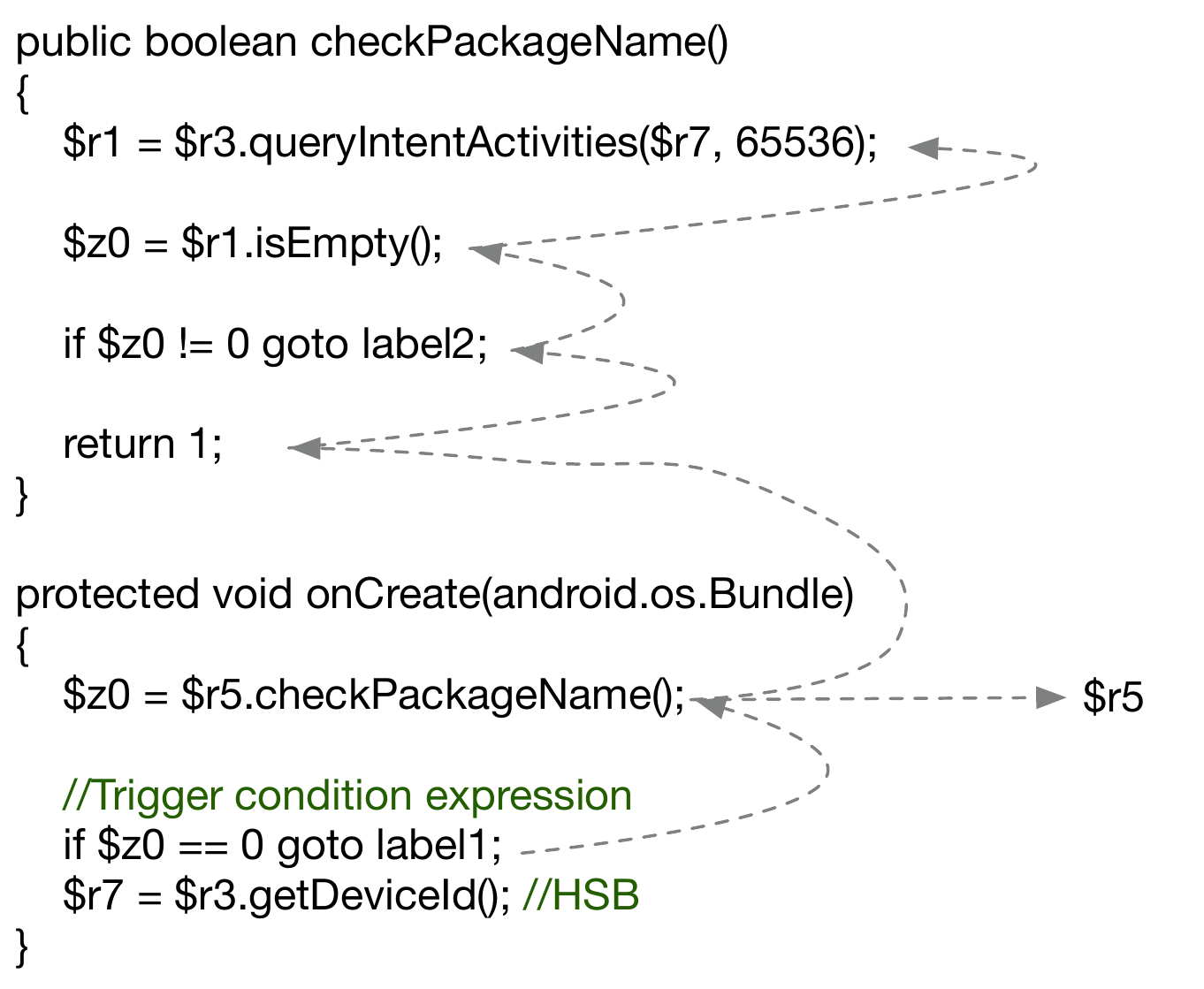}
	\caption{The simplified working process of the trigger condition inference module. The code is presented in simplified Jimple, which is an intermediate representation of Soot~\cite{lam2011soot}. Soot is the underline static analysis framework leveraged by \tool{} to achieve the backward data-flow analysis.}
    \label{fig:triggercondition}
\end{figure}

When inferring the \emph{definition statement}, inter-procedural analysis needs to be taken into account because the trigger conditions can be defined in other methods and transferred to the HSB via callee's returned values or caller's parameter values.
Indeed, take the code snippet shown in Listing~\ref{code:emu} as an example,  the trigger condition is actually defined in method \emph{checkPackageName()} despite the HSB is seated in the \emph{onCreate()} method.
Fig.~\ref{fig:triggercondition} illustrates the backward tracking flow showing how our approach identifies the trigger condition.
When there is a method involved in the backward tracking flow, our data-flow analysis will keep tracking the method's caller object as it may be relevant to the definition of the trigger condition.
For example, our analysis will keep tracking \$r1 when statement \emph{\$r1.isEmpty()} is reached.
If the method is a user-defined function, our data-flow analysis will further jump into the method and keep tracking its returned variables (all variables will be tracked if there are several return statements).
The backward data-flow analysis will terminate if System APIs are identified, or Android's entry-point methods (such as UI callback methods or components' lifecycle methods) are reached.
The analysis will also stop if the condition is linked to a constant value that is further not originated by \emph{if-statements}.

\subsection{Suspicious HSO Detection}
\label{sec:common}

The last module takes the outputs of the previous module to detect hidden sensitive operations, following the rules presented in Definition 1 and Definition 2 (cf. Section~\ref{sec:motivation}).
Unfortunately, these rules are not perfect and 
may introduce false-positive results that have similar characteristics of HSOs but are actually user intended behaviors. Indeed, for the same operations, under different circumstances, they could be flagged as conventional usages or suspicious operations and could lead to benign or user intended malicious behaviors.
These false results include common programming patterns used in legitimate \emph{if-then-else statements} (hereinafter referred to as \emph{conventional usages}), which should be excluded by \tool{} . Therefore, we resort to building a list of conventional usages (or whitelist) and based on it, in the last module of \tool{}, we filter out non-malicious HSOs and only keep suspicious HSOs. 

Nevertheless, we argue that it is non-trivial to understand the developer's intention behind the operations.
Therefore, in this last module, in addition to automatically detect suspicious hidden sensitive operations, \tool{} goes one step deeper to also provide adequate details to explain why an suspicious HSO is flagged as such, i.e., what is the trigger condition, what is the logic of the \emph{if condition}, and what are the sensitive behaviors triggered if the logic is fulfilled.
This function is provided for helping security analysts understand whether the flagged HSOs should be regarded as malicious or not.

By leveraging \tool{}, in Section~\ref{sec:commonHSO}, we study and collect \emph{conventional usages} in large sets of Android apps, whereas in Section~\ref{sec:evaluation}, we put the emphasize on \emph{suspicious} HSOs.

\section{Conventional Usage Analysis}
\label{sec:commonHSO}

The overall goal of this work is to detect hidden sensitive operations so as to unveil the evasive technologies that are frequently leveraged to hide malicious behaviors. 
In this section, we evaluate our approach based on a large set of Android apps towards checking if our approach \tool{} is capable of fulfilling this goal.
Specifically, in this section, we conduct an exploratory study of recent hidden sensitive operations aiming to understand the current status quo of conventional usages and build a comprehensive list of conventional usages (to be used by \tool{} to discriminate suspicious HSOs from conventional usages).

Recall that our approach, in its last working step, takes as input a customizable list of conventional usages to filter out non-suspicious HSOs, which subsequently helps in saving significant security analysts' efforts as they now only need to scrutinize the retained small number of likely suspicious HSOs.
Towards identifying such conventional usages, we apply a semi-automatic process to summarize based on their frequency of occurrence. The conventional usage whitelist is built based on reasonable assumptions that legitimate HSOs frequently appear in Android apps, including both malware\footnote{Malware is included because often not all of its code is malicious. It might contain a malicious payload but the other code could still remain benign.} and goodware.
We manually inspected the trigger APIs that have appeared more than 50 times in our dataset and determined if they should be categorized as a conventional usage. By doing so, we defined seven major categories of conventional usages. Also, the results of our manual analysis are cross-validated by two authors. The two authors first independently conduct the manual analysis (to discover knowledge with support evidence from various software artifacts). They then had physical meetings to discuss, merge, and finalize the results.

\textbf{\emph{Experimental\ Setup}}.
We applied \tool{} (with the list of conventional usages set to be empty\footnote{The experimental results should contain both conventional usages and suspicious HSOs.}) on a dataset that contains 10,000 malware samples (referred to as \textit{malware set}) and 10,000 benign apps (referred to as \textit{benign set}).
The \textit{malware set} was collected from VirusShare \cite{Virusshare} from 2012 to 2020. To better reflecting recent trends on the deceptive techniques used in malware samples, we only include the samples whose first seen date was on or after 2016. The malware samples were submitted to VirusTotal\footnote{https://www.virustotal.com} for screening, and only the ones that have been labelled by more than five anti-virus engines (VirusTotal has hosted over 70 anti-virus scanners) were selected.

The \textit{benign set} was randomly selected from a pool of more than 100,000 apps crawled from Google Play in 2019, which are further scanned to ensure non of them are tagged by VirusTotal.

Our tool has identified 45,342 HSOs (35,974 in the \textit{malware set}, and 9,368 in the \textit{benign set}) triggered by 54,152 conditions. Note that some HSOs may be triggered by more than one condition (e.g., multiple conditions in a CTB that are connected by \emph{AND} or \emph{OR} operators). 
Towards evaluating the precision of \tool{}, i.e., the identified HSOs meet our previous rule definitions, we manually examine 20 randomly selected APKs from the total 8,107 apps that have been identified to contain at least one HSO. From these apps, our approach identified 157(with a confidence level of 95\% and a confidence interval of 7.81\%) HSOs in total, among which 155 of them are eventually confirmed to be true HSOs, giving an precision of 98.7\%. This result suggests that \tool{} is capable of identifying HSOs in Android APIs.

\begin{figure}[!t]
    \centering 
    \includegraphics[width=0.65\textwidth]{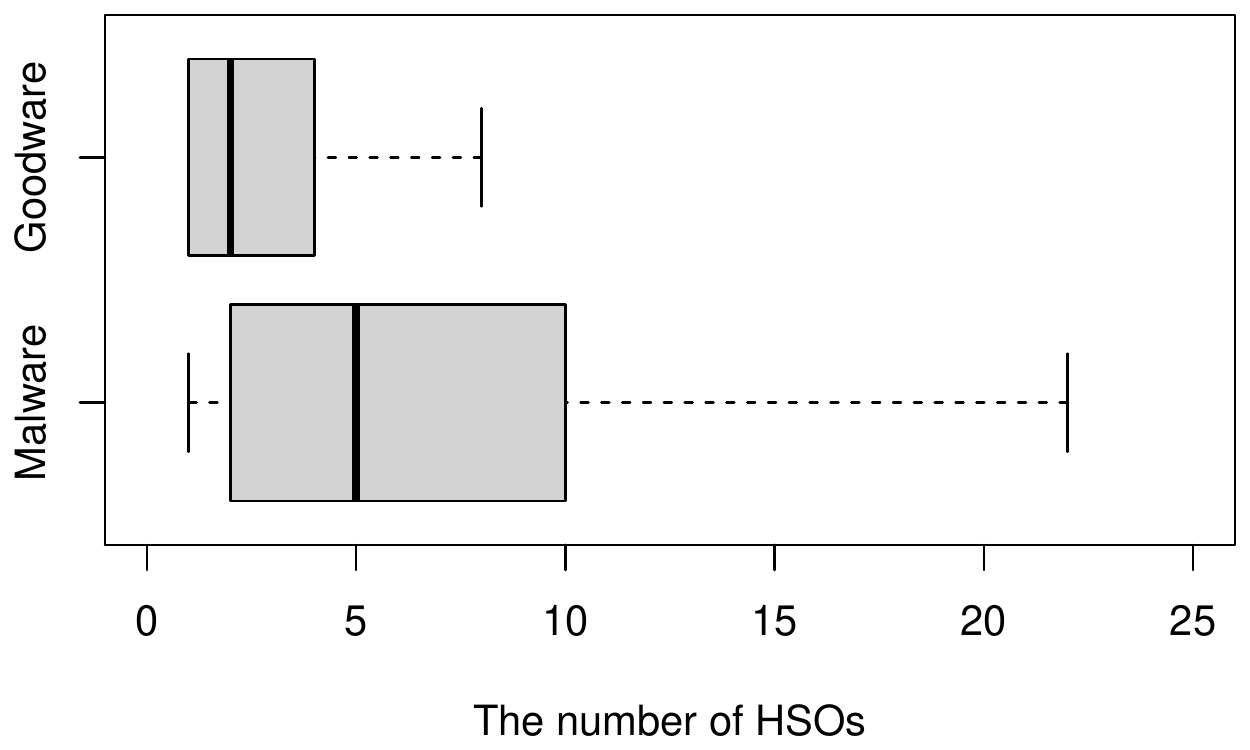}
    \caption{Distribution of the number of HSOs in \textit{benign set} and \textit{malware set}.}
    \label{fig:HSO_common_cases}
\end{figure}

Figure~\ref{fig:HSO_common_cases} further presents the distribution of the number of HSOs detected in the apps from \textit{benign set} and \textit{malware set}.
Expectedly, malware samples involve significantly more HSOs than that of benign apps, as confirmed by the \emph{p-value} a Mann-Whitney-Wilcoxon (MWW) test at a significance level\footnote{Given a significance level $\alpha = 0.001$, if p-value $< \alpha$, there is one chance in a thousand that the difference between the two datasets is sue to a coincidence.} at 0.001~\cite{fay2010wilcoxon}.
This result suggests that HSOs are more favored by malware than benign apps.
Hence, our community should pay more attention to the appearance of HSOs to help security analysts better dissect malicious apps.

Based on the previous experimental results, we manually analyzed the trigger conditions and the corresponding hidden operations to identify \emph{conventional usages}, i.e., HSOs (at least based on our definition) that are actually legitimate and occur relatively often in Android apps. We first inspected the trigger APIs that have appeared more than 50 times in our dataset and determined if it is a \emph{conventional usage}. By doing so, we identified seven major categories of \emph{conventional usages}. Then we reviewed each of the rest of the cases to further filter out the other \emph{conventional usages}.
Finally, 43,141 \emph{conventional usages} have been identified, out of which 40,412 cases belong to the seven major categories.
As the whitelist is generated by manual analysis, it cannot cover all possible \emph{conventional usages}. However, we believe that the majority of the \emph{conventional usages} have been identified (i.e., from the seven categories), new special cases can always be added to the list and incorporated into \tool{} in the future. 
We now elaborate on the seven major categories, each with an example code snippet presented in Listing \ref{code:commoncases}.

\begin{lstlisting}[
caption={Examples of conventional usages.},
label=code:commoncases,
firstnumber=1,float=t]
// conventional usages - SDK version check
public void addJavascriptInterface(Object var1, String var2) {
 if (VERSION.SDK_INT < 17) {
  TaoLog.e("HybridWebView", "addJavascriptInterface is disabled before API level 17 for security.");
 } else {
  super.addJavascriptInterface(var1, var2);
}}

// conventional usages - User Interface
class ClickEvent implements View.OnClickListener {
    public void onClick(View view) {
        if(view.getId() == backButton.getId()){
            webView.goBack()
        }
        else if (view.getId() == reloadButton.getId()){
            webView.reload();
}}}

// conventional usages - File Handling
public static File inputstreamtofile(InputStream ins) {
 File SDFile = Environment.getExternalStorageDirectory();
 File desDir=new File(SDFile.getAbsolutePath());
 File newFile=new File(desDir.getAbsolutePath() + File.separatorChar+"myPaint.png") ;
 if(desDir.exists()){
  OutputStream os = new FileOutputStream(newFile);
  while ((bytesRead = ins.read(buffer, 0, 8192)) != -1) {
   os.write(buffer, 0, bytesRead);
}}}

// conventional usages - Permission Check
public static String getDeviceInfo(Context context) {
 if (checkPermission(context, Manifest.permission.READ_PHONE_STATE)) {
  String device_id = tm.getDeviceId();
 } else {
  requestPermissions(context, new String[] {Manifest.permission.READ_PHONE_STATE}, REQUEST_CODE)
}}

// conventional usages - Network
public String g() {
var1 = ((ConnectivityManager)a.getSystemService("connectivity")) .getActiveNetworkInfo();
var9 = var1.getType();
if(var9 == 1){
 var10 = ((WifiManager)a.getSystemService("wifi")).getDhcpInfo();
}}

// conventional usages - Intent Management
public void onReceive(final Context context, Intent intent) {
 String action = intent.getAction();
 if (action.equalsIgnoreCase ("android.net.conn.CONNECTIVITY_CHANGE") {
  connectivityManager.getActiveNetworkInfo();
}}

// conventional usages - SharedPreferences
public class VpnAddressIp{
 public SharedPreferences sp;
 public String VPNAddress() {
  sp = context.getSharedPreferences("SP", Context.MODE_PRIVATE);
  VPNflag = sp.getInt("VPNFlag", 1);
  VPNAddress vpnaddress = new VPNAddress(context);
  if (VPNflag == 1) {
   VPNAddressBean bean = vpnaddress.queryVPN(1);
   networkaddress = bean.getNetwork();
  } 
  return networkaddress;
}}
\end{lstlisting}

\textbf{\emph{SDK Version.}}
With Android system update, new APIs are defined to replace old ones. To maintain the compatibility of apps across different Android versions, it is a common practice to check the SDK version before deciding the right API to use. 
Lines 2\textasciitilde7 show a simplified code snippet of a legitimate conventional usage that fulfills all the rules we defined for an HSO. The code checks whether the Android version is newer than \emph{Android level 17} (line 3), if so, the app leverages the \emph{addJavascriptInterface()} API to inject Javascript into the \emph{WebView} (line 6), otherwise, it logs an error message (line 4) as the API is not available in the Android version lower than 17. While SDK version check commonly exists in both malware and benign apps (with 10,303 and 3,223 cases, respectively), this check does not intend to hide the behaviors within the \emph{if-then-else} statement and hence should be excluded from the HSO results.

\textbf{\emph{User Interface.}}
When the user interacts with UI widgets (e.g., press a button), it retrieves and compares the UI widget's \emph{id} (i.e., a system API) to determine which widget has been fired. If there happened to be a sensitive API invoked in one of the branches' statements, this code block will be misidentified as HSO. Lines 10\textasciitilde17 show an example of a button's callback method, which checks the ID of the buttons (lines 12,15) and either go back to the previous webpage (line 13) or reload the current page (line 16). User Interface has 8,052 instances in the \textit{malware set} and 2,426 instances in the \textit{benign set}. 

\textbf{\emph{File.}}
The existence of a file or a directory is usually checked before file operations, such as reading and writing files. The code for checking file existence typically put the subsequent actions in one branch (where the file does exist), and show an error message in the other branch (where the file does not exist). In some cases, it even has only the \emph{if-branch}. Therefore, it satisfies the rules mentioned above and will be mistakenly identified as an HSO. Our results have observed 7,217 and 1,701 cases in our \textit{malware set} and \textit{benign set}, respectively.
A file checking example can be found in lines 20\textasciitilde28, where the code checks the existence of an external storage (line 24), and copy an image there (lines 26,27). 

\textbf{\emph{Permission.}}
Since Android 6.0, the dangerous-level permissions need to be explicitly checked and requested before accessing the APIs protected by these permissions. The example code for checking permission can be found in lines 31\textasciitilde36. It first checks whether the app has been granted \emph{READ\_PHONE\_STATE} permission (line 32). Then, the app either invoke the permission protected API (line 33) or request the missing permission (line 35) based on the check result. Even though a sensitive API \emph{getDeviceId()} is called in one branch, which behaves quite differently than the other branch, it does not mean to hide this behavior. Therefore, it is regarded as a \emph{conventional usage}. Permission check has appeared 6,727 and 936 times in the HSOs identified in the \textit{malware set} and \textit{benign set}, respectively.


\textbf{\emph{Network.}} Network information (e.g., network type, connection status, etc.) is always checked before performing network-related behaviors, ensuring that the network status is suitable for accomplishing the subsequent tasks. For example, the network type (e.g., WiFi, cellular, etc.) is checked before downloading large files, and if it is on the cellular network, the download will be suspended. Another example demonstrated in lines 39\textasciitilde44 examines the type of connected network (line 41) and get its DHCP information if the phone is connected to WiFi (1 is the value of ConnectivityManager\#TYPE\_WIFI). There are 5,224 an 744 identified HSO cases that are related to the \emph{Network} in the \textit{malware set} and the \textit{benign set}, respectively.


\textbf{\emph{Intent.}}
\emph{Intent} is a crucial mechanism to assist the communication between different components in the Android system. \emph{Intent} has various legitimate usages, including starting activities and services, passing data and properties, etc. Lines 47\textasciitilde51 demonstrate a legitimate example of handling the callback method of receiving an \emph{Intent}. In this example, it checks the \emph{action} defined in the received \emph{Intent}, and calls \emph{getActiveNetworkInfo()} method (i.e., a sensitive API) if the \emph{action} is \emph{CONNECTIVITY\_CHANGE}. There are 1,911 an 1,011 identified HSO cases that are related to the \emph{Intent} in the \textit{malware set} and the \textit{benign set}, respectively.


\textbf{\emph{SharedPreferences.}}
In Android system, data can be saved as <key, value> pairs and stores as a \emph{SharedPreferences} object in a file that can be accessed by \emph{getSharedPreferences()} interface. It provides a lightweight and easy-access data store mechanism, which is widely used in storing small collection of data, such as configurations of the app. Reading the values from the \emph{SharedPreferences} and action accordingly is considered a legitimate behavior.  
Lines 54\textasciitilde65 illustrate an example of using \emph{SharedPreferences}, where the code retrieves the value of a configuration item ``VPNFlag'' (line 58) from a \emph{SharedPreferences} object named ``SP'' (line 57), and query the corresponding VPN services accordingly (lines 60\textasciitilde62). There are 878 and 358 cases involving the usage of \emph{SharedPreferences} in our \textit{malware set} and \textit{benign set}, respectively. 

\textbf{Completeness of conventional usages.}
\label{subsec:Completeness_conventional_usages}
Since the conventional usage categories are summarized with manual efforts on a given set of apps, they may not be representative and thereby may not cover all possible cases.
Therefore, in this work, we go one step deeper to further investigate the completeness of all the seven categories of conventional usages by applying our approach to another set of randomly selected 10,000 malware and 10,000 benign apps from AndroZoo~\cite{liu2020androzooopen}. 
We remind the readers that AndroZoo includes over 10 million Android apps that were collected from both the official Google Play store and several third-party app markets. To avoid potential biases in our results, we made additional efforts to remove potentially duplicated apps (i.e., different versions of the same app), and only the latest version is retained. For the 20,000 apps, we apply \tool{} to analyze these apps and inspect the trigger conditions that have appeared more than 50 times. We then manually determine if they are conventional usage. To do this, two of the authors spent ten person-days manually summarizing conventional usages (e.g., API-API or Key-API pairs). After manually checking the experimental results and the bytecode of apps, we have totally picked up 41,035 conventional usages, among which 38,920 cases fall into the predefined whitelist (with a success rate of 94.8\%). This result shows that, despite testing on different apps, our whitelist is still quite stable and effective in eliminating conventional usages.

Apart from the aforementioned commonly appeared conventional usages, we further look into some of the uncommon conventional usages. Our manual observation confirms that those uncommon conventional usages are indeed legitimate HSOs that do not appear frequently in Android apps. We present two concrete examples of uncommon conventional usages to illustrate this concept. One example is that an app first checks if the directory of downloads exists (i.e., a standard directory to place files that have been downloaded by the user), and then automatically starts the download using the \emph{android.app.DownloadManager\#enqueue} API once the download manager is ready and connectivity is available. We consider it a conventional usage because it is against the second principle of suspicious HSO's definition: the user does not intend to hide such behavior. In addition, given that there exist several substitute ways of downloading files (e.g., Http request, URLConnection, BufferedInputStream, FileOutputStream, etc.), the native APIs lie in \emph{android.app.DownloadManager} are not that commonly used by app developers. Thus, we regarded it as an uncommon conventional usage. As another example, the sensitive behavior of vibration could be triggered only when a user clicks a certain button. We consider it also a conventional usage because it involves non-hidden behaviors. In fact, if app developers intend to hide sensitive behaviors, it would be obvious that they won't use vibration functionality to notify users. Moreover, the usage of vibration is less common because it would annoy Android users, leading to a poor user experience. Such cases appear less than 50 times in our dataset and thus we regard them as uncommon conventional usage as well.

\section{Suspicious HSO Analysis}
\label{sec:evaluation}

\begin{figure}[h!]
    \centering 
    \includegraphics[width=0.6\textwidth]{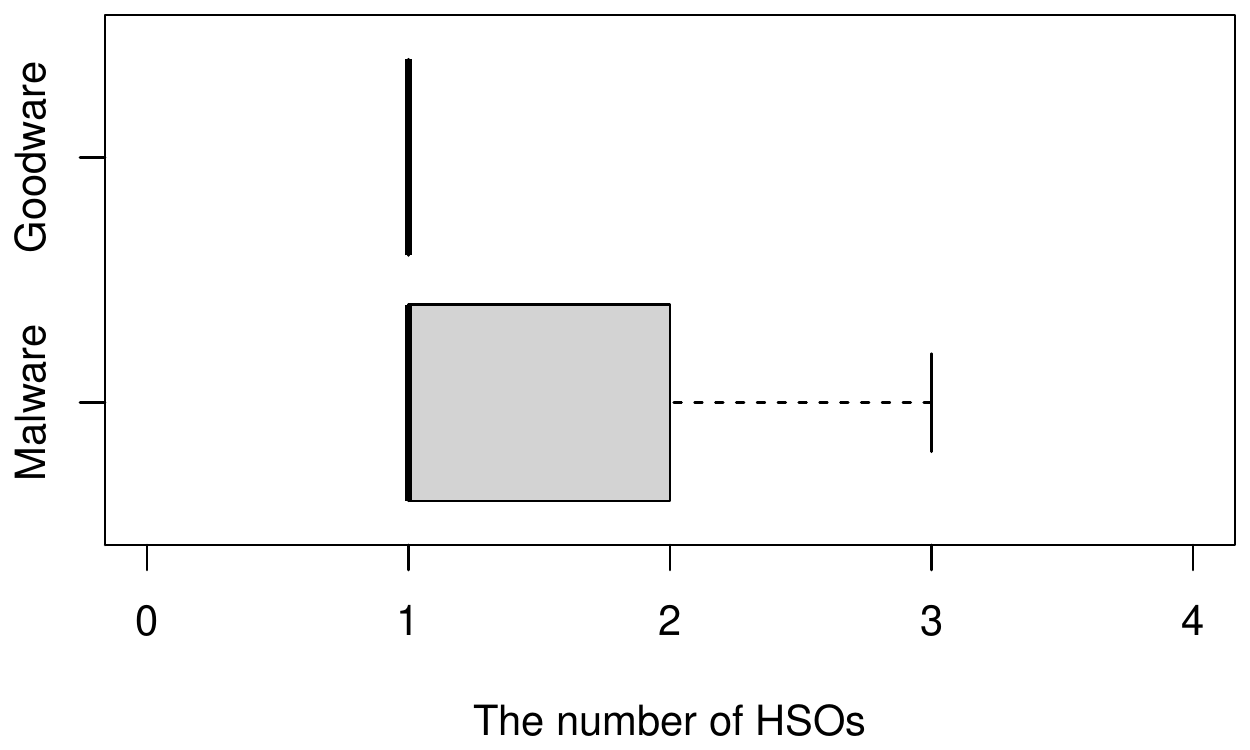}
    \caption{Distribution of the number of \textit{suspicious} HSOs in \textit{benign set} and \textit{malware set}.}
    \label{fig:HSO_no_common_cases}
\end{figure}

After eliminating conventional usages, all the remaining ones will be reported as suspicious HSOs.
Among the 20,000 apps considered in this work, 1,304 of them, including 982 malware and 322 benign samples, were retained.
These apps have been reported to contain in total 2,201 suspicious HSOs, with 1,790 and 441 from malware and benign apps, respectively.
These numbers are recapped in Table~{\ref{tab:numberOfSuspiciousHSOs}}. 
This experimental result shows that suspicious HSOs are widely present in real-world Android apps. 
Figure {\ref{fig:HSO_no_common_cases}} further illustrates the distribution of suspicious HSOs in our dataset. On average, there are 2.0 and 1.4 HSOs in each malware sample and benign app, respectively.

\begin{table}[!h]
\centering
\caption{Number of suspicious HSOs.} 
\label{tab:numberOfSuspiciousHSOs}
\resizebox{0.6\linewidth}{!}{
\begin{tabular}{r| l l } 
Initial Dataset   & \# HSOs & \# Suspicious HSOs \\

\hline
\hline
10,000 benign apps  & 9,368 (in 3,071 apps)  & 441 (in 322 apps)\\

\hline
10,000 malicious apps & 35,974 (in 5,036 apps) &  1,790 (in 982 apps)\\ 

\hline
Total & 45,342 (in 8,107 apps)  &  2,201 (in 1,304 apps)\\ 
\end{tabular} 
}
\end{table}

In this work, the elimination of conventional usages is based on a pre-defined whitelist, which only includes recurrently presented HSOs in benign apps. 
Some less frequent yet still legitimated HSOs could have been overlooked and hence result in suspicious ones.
Indeed, the remaining suspicious HSOs may not always be true positives (i.e., may contain a small number of false positives).
To this end, we go one step further to calculate the precision of our approach in pinpointing suspicious HSOs in Android apps.
Unfortunately, there is no known ground truth available for evaluating HSO usage in Android apps.
Thus, we resort to a manual process to calculate the precision. In this work, we manually inspected the bytecode of apps to see if \tool{} correctly and precisely identified the suspicious trigger rather than those commonly appeared code blocks for normal usage. Here, we identify truly suspicious behavior (i.e., confirmed to be true positive) only when the HSO is security-relevant and potentially brings harm to Android users.
Specifically, we rely on two principles to identify truly suspicious HSOs: (1) the hidden behavior involves security-relevant APIs that are protected by Android permissions, classified following the latest Android API-permission mappings (cf. Section 3.1), and (2) the sensitive APIs are intentionally hidden under dedicated trigger conditions. As a result, we count those who meet the two aforementioned principles as true positives. For example, if an app first intends to retrieve Device ID, and when unsuccessful, tries to read the MAC address, we will consider it as a false positive because it is against the second principle: does not intend to hide such behavior.  As another example, an app checks the build's fingerprint to see if it is running on popular emulators, and the sensitive behavior of retrieving subscriberId would be triggered only when it is not running in an emulator. We consider it as a true positive because it involves security-relevant APIs and there is sensitive behavior that is clearly hidden under trigger conditions.
In our dataset, 1,304 apps have been reported to contain at least one HSO. Among the 1,304 apps, HiSenDroid has identified 14,394 HSOs, for which 2,231 of them are regarded as suspicious HSOs. By manually looking at each of those reported suspicious HSOs, we are able to confirm that 1,938 out of 2,231 of them are true positive results (or 293 of them cannot be confirmed without deeply examining the code), giving a precision of 86.8

Recall that the conventional usages are excluded in this work through a whitelist built through empirical evidence, and the whitelist is only considered as a configuration option to our approach.
We believe that the performance of detecting suspicious HSOs could be further improved if we are able to construct a better whitelist of legitimate HSOs. This is nevertheless outside the scope of this work. We hence consider it as our future work.

\subsection{Trigger Conditions}

Known trigger types such as time-bomb and anti-emulator techniques have been broadly studied, specific algorithms for detecting such known trigger types have been developed~\cite{balzarotti2010efficient, kirat2015malgene, fratantonio2016triggerscope}. Nevertheless, the community still lacks the understanding of unknown trigger types. We therefore investigate the most frequent triggering conditions in the suspicious HSOs detected by \tool{}. 

\tool{} has identified 168 unique APIs that have been traced as the source of the triggers in detected suspicious HSOs. To make it much clearer, we present all these system properties trigger conditions and environment parameters trigger conditions in the artefact package ~\footnote{https://bitbucket.org/se\_anonymous/hisendroid/src/master/experiments\_results/}. We then manually categorize them according to the types of objects they accessed. Table~\ref{tab:top10trigger} illustrates the trigger condition categories and examples of system APIs that are frequently leveraged to fulfill the trigger conditions discovered in HSOs.
The top trigger condition categories include time (e.g., at a certain time of a day), SMS (e.g., when receives SMS of certain formats), location (e.g., if the device is in certain countries), 
system property (e.g., checks the device's manufacturer), and package manager (e.g., if specific apps are installed).

\begin{table}[!h]
\centering
\caption{Categories of Trigger Conditions in HSO} 
\label{tab:top10trigger}
\resizebox{0.65\linewidth}{!}{
\begin{tabular}{| r | l |} 
\hline
Category &  Most Frequent Trigger API Examples \\
 \hline
\multirow{3}{*}{Time}       & 
 util.Calendar\#get    \\
 &  util.Date\#getTime  \\
 &  util.Calendar\#getTimeInMillis\\
\hline
\multirow{3}{*}{System Properties}     & 
 os.Build\#MODEL \\
&  telephony.TelephonyManager\#getSubscriberId   \\
&  telephony.TelephonyManager\#getDeviceId  \\

\hline
\multirow{3}{*}{Location}       & 
 telephony.TelephonyManager\#getSimCountryIso \\
 &  telephony.TelephonyManager\#getCellLocation \\
 &  location.LocationManager\#getLastKnownLocation \\
\hline
\multirow{3}{*}{SMS Message}     & 
telephony.SmsManager\#divideMessage \\
 &  telephony.SmsManager\#getDefault \\
 &  telephony.SmsManager\#getData \\  
\hline
\multirow{3}{*}{Package Manager}     &  
 content.Context\#getPackageManager \\
 &  content.pm.PackageManager\#getPackageInfo  \\
 &  content.pm.PackageManager\#getApplicationInfo \\
\hline
\multirow{3}{*}{Miscellaneous}     & 
 android.widget.CheckBox\#isChecked \\
 & android.app.KeyguardManager\#isKeyguardLocked  \\
 &  java.net.NetworkInterface\#getHardwareAddress \\
\hline

\end{tabular} 
}
\end{table}

Here we elaborate on each trigger condition category with real-world suspicious HSO cases identified in our dataset. 

\textbf{\emph{Time Triggers}} compare time-related properties (such as current system time, time zone, etc.) with hard-coded values to determine whether or not to execute the hidden sensitive behaviors. Listing \ref{code:Time} demonstrates a code snippet from app \emph{com.wukongtv.wukongtv}\footnote{SHA-256:3397079daa388bdbcdcc42b6834d3c792bf5c80ad24491e3893de7cfc2b11db7}, which leverages time-related triggers to hide suspicious behaviors. When the first time the app launches, it writes the timestamp into the SharedPreferences (i.e., \textit{var0}). It then compares the current system time with the first launch time (line 6); if the time interval is greater than one day, it triggers the sensitive method \textit{bq.f()} (line 7) that retrieves the information (e.g., package name and process name) of running tasks (lines 10\textasciitilde15). Doing so conceals the suspicious behaviors from automatic dynamic detection, which usually starts testing immediately after the app is installed.  

\begin{lstlisting}[style=JAVA, escapechar=\%,
caption={Code Example of Time Trigger.},
label=code:Time,
firstnumber=1]
static void c(Context var0) {
 Calendar var10000 = Calendar.getInstance();
 int var2 = var10000.get(6) * 100;  //day_of_year
 var2 += var10000.get(11);          //hours_of_day
 // var0 is retrieved from SharedPreferences
 if %\underline{(Math.abs(var0/100L - (long) (var2/100)) >= 1L)}% {
  ab.h = bq.f(var0);
}}}

public static Long[][] f(Context var0) {
 var31 = var3.getRecentTasks(10, 1);
 while(var31.iterator().hasNext()){
  // get package name and process name of recent tasks
  ...
}}
\end{lstlisting}

\textbf{\emph{System Property Triggers}} leverage system properties, such as the phone model, the phone number, and hardware information, to limit the sensitive behaviors within specific device brands (e.g., Samsung) or types (e.g., real device). These triggers are also commonly adopted by anti-emulator techniques to detect the presence of emulators. Listing \ref{code:System_Properties} demonstrates an anti-emulator example extracted from app \emph{com.gwsoft.imusic.controller} \footnote{SHA-256:8c679a7c57a7fbb355fb363d3784cc8380655701d482837869edd95f3a3ea470}, which checks if the build's fingerprint contains specific strings that indicate popular emulators (line 2). The sensitive behavior of retrieving \textit{subscriberId} (line 5) is only executed if it does not run in an emulator. 

\begin{lstlisting}[style=JAVA, escapechar=\%,
caption={Code Example of System Property Trigger.},
label=code:System_Properties,
firstnumber=1]
private static boolean a(Context var0) {
 if %\underline{(Build.FINGERPRINT.contains("vbox86p/vbox86p")\&\&} \underline{!Build.FINGERPRINT.contains("ttVM\_Hdragon/ttVM\_Hdragon")\&\&} \underline{!Build.FINGERPRINT.contains("generic/sdk/generic")\&\&} \underline{!Build.FINGERPRINT.contains("generic\_x86/sdk\_x86/generic\_x86")}%
 // omit other strings that fingerprints popular Android emulators
 ){
  var2 = ((TelephonyManager)var0.getSystemService("phone")) .getSubscriberId();
  var11.put("imsi", var2);}}
\end{lstlisting}

\textbf{\emph{SMS Triggers}} Utilize the content, type, and phone number of received SMS messages to hide sensitive behaviors. An example derived from app \emph{com.fingersoft.hillmotor}\footnote{SHA-256:95e1cf498dec79351a9d104f5e9fb0110c267e9eff0099ada475d8832a2afb7302521} is shown in Listing \ref{code:SMS_Message}. When an SMS message is received, it checks the originating address of the message (line 4). If it matches a pre-defined value (e.g., 10 or 11 etc in this example), the behavior that repeatedly sends a message (line 6) to the same number via a text message service.

\begin{lstlisting}[style=JAVA, escapechar=\%,
caption={Code Example of SMS Trigger.},
label=code:SMS_Message,
firstnumber=1]
//var2 is the originating address retrieved from SMS
//var3 is the message body
public boolean repeat(Context var1, String var2, String var3) {
if ((var2.startsWith("10") || var2.startsWith("11") || var2.startsWith("12")) && !var2.equals("114") && !var2.equals("12306") && !var2.equals("116114") && !var2.equals("12580")) {
SmsManager var13 = SmsManager.getDefault();
var25.sendTextMessage(var2, (String)null, "Y", var16, var12);
}}
\end{lstlisting}

\textbf{\emph{Location Triggers}} obscure sensitive behaviors with fine grained (e.g., latitude and longitude) and coarse grained (e.g. country) location information. Listing \ref{code:Location} shows an example derived from \textit{com.inter.\\apps.patqut.apk} \footnote{SHA-256:22c9d7738073a7ac8f9b58029057c2741e89faac76b623837db2f3a8bb2d93c5} which queries the country code of the device (saved as \textit{var1}), and checks if it is in Malaysia (line 4). If so, the app then triggers the \textit{postLoginData2()} method (line 5), which retrieves the device's id (line 9) and hands it over to another activity for further malicious behaviors. 

\begin{lstlisting}[style=JAVA, escapechar=\%,
caption={Code Example of Location Trigger.},
label=code:Location,
firstnumber=1]
TelephonyManager var3 = (TelephonyManager)this.getSystemService("phone");
String var1 = var3.getSimCountryIso().toUpperCase();
public void getin(String var1) {
 if %\underline{(var1.equals("MY"))}% {
  this.postLoginData2();
}}

public void postLoginData2() {
 String var2 = ((TelephonyManager)this.getSystemService("phone")) .getDeviceId();
 // hand over the obtained DeviceId to a new activity
 ...
}
\end{lstlisting}

\textbf{\emph{Package Manager Triggers}} scan the list of installed apps and inspect if specific apps (usually anti-virus tools) are installed before conducting any sensitive behaviors.
Listing~\ref{code:Check_PackageName} shows a code snippet taken from \emph{flash15.1.apk} \footnote{SHA-256:fdaba7f032ee7ff9adf799713b25d4c2fef86ddbbe8709bf6ec021505b8f1d0d} which searches for \emph{AhnLab V3 Mobile Plus 2.0} (i.e., an anti-virus tool) in the list of installed apps (line 1\textasciitilde7). If the anti-virus tool is not installed (line 10), it then starts its malicious behaviors. Specifically, it gets the package name of the current active activity (lines 11, 12), and puts it into sleep if it is a bank app. After that, it launches a new activity that contains a phishing web page to steal user's bank credentials.

\begin{lstlisting}[style=JAVA, escapechar=\%,
caption={Code Example of Package Manager Trigger.},
label=code:Check_PackageName,
firstnumber=1]
private boolean judgeAV() {
 this.pm = this.getPackageManager();
 this.listAppcations = this.pm.getInstalledApplications(8192);
 for(int v = 0; v < listAppcations.size(); ++v) {
  if(listAppcations(v).name.equalsIgnoreCase("AhnLab V3 Mobile Plus 2.0")){
   return true;}
  return false;}
    
public void run() {
if %\underline{(!AutBan.this.judgeAV())}% {
 List var2 = ((ActivityManager)AutBan.this.getSystemService ("activity")).getRunningTasks(1);
 String var1 = ((RunningTaskInfo)var2.get(0)).topActivity .getPackageName();
 // if the top activity is a bank app, it puts the activity into sleep and start a phising page
 ...
}};
\end{lstlisting}

\textbf{\emph{Other Triggers.}} Besides the most frequent trigger categories, we also observed some sophisticated triggers specifically designed to counter automated dynamic testing approaches. 
Listing~\ref{code:Item_Click} shows an example taken from a music player app \emph{com.gwsoft.imusic.controller}\footnote{SHA-256:8c679a7c57a7fbb355fb363d3784cc8380655701d482837869edd95f3a3ea470}. The app hides sensitive behaviors that retrieve the device's information (lines 8\textasciitilde12) behind a trigger that will only be fired when an item on the song list (i.e., \textit{mCatalogSongsList}) is clicked (line 2). The trick here is that automated dynamic testing tools running on an emulator are likely not to have any music files and, therefore, will have no items on the list to click. Hence, only legitimate users who intend to use it to play music will have the chance to trigger the sensitive behavior. 

\begin{lstlisting}[style=JAVA, escapechar=\%,
caption={Code Example of Other Trigger.},
label=code:Item_Click,
firstnumber=1]
//contains at least one song in the list
public void onItemClick(AdapterView<?> var1, View var2, int var3, long var4) {
 if %\underline{(mCatalogSongsList != null \&\& var3 + -1 >= 0 \&\& var3 + -1} \underline{< mCatalogSongsList.size())}%{
  CountlyAgent.onEvent(CuttingActivity.this,  "activity_diy_do_re", String.valueOf(var3));
}}

public static void onEvent(Context var0, String var1, String var2) {
 HashMap var3 = new HashMap;
 var3.put("phone", getIMSI());
 var3.put("ip", getLocalIpAddress());
 var3.put("app_version", versionName);
 var3.put("imei",getDeviceId());
}
\end{lstlisting}



\begin{figure}[t!]
    \centering
    \subfigure[Categories of Trigger Conditions in HSO.]{\label{fig:top10trigger}\includegraphics[width=0.45\linewidth]{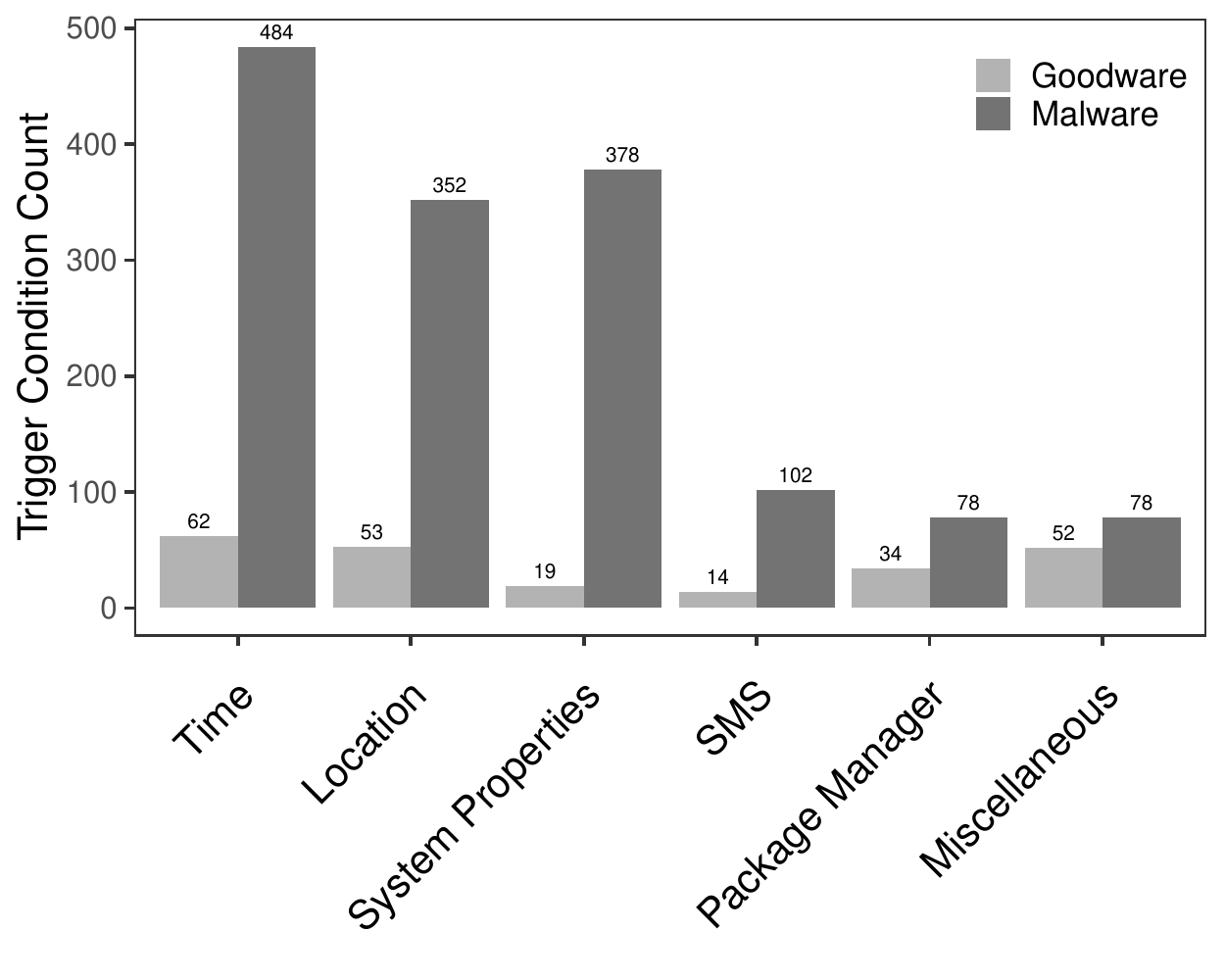}}
    \subfigure[Top 10 Categories of Sensitive APIs in Malware.]{\label{fig:malware_goodware_sensitive_api}\includegraphics[width=0.45\linewidth]{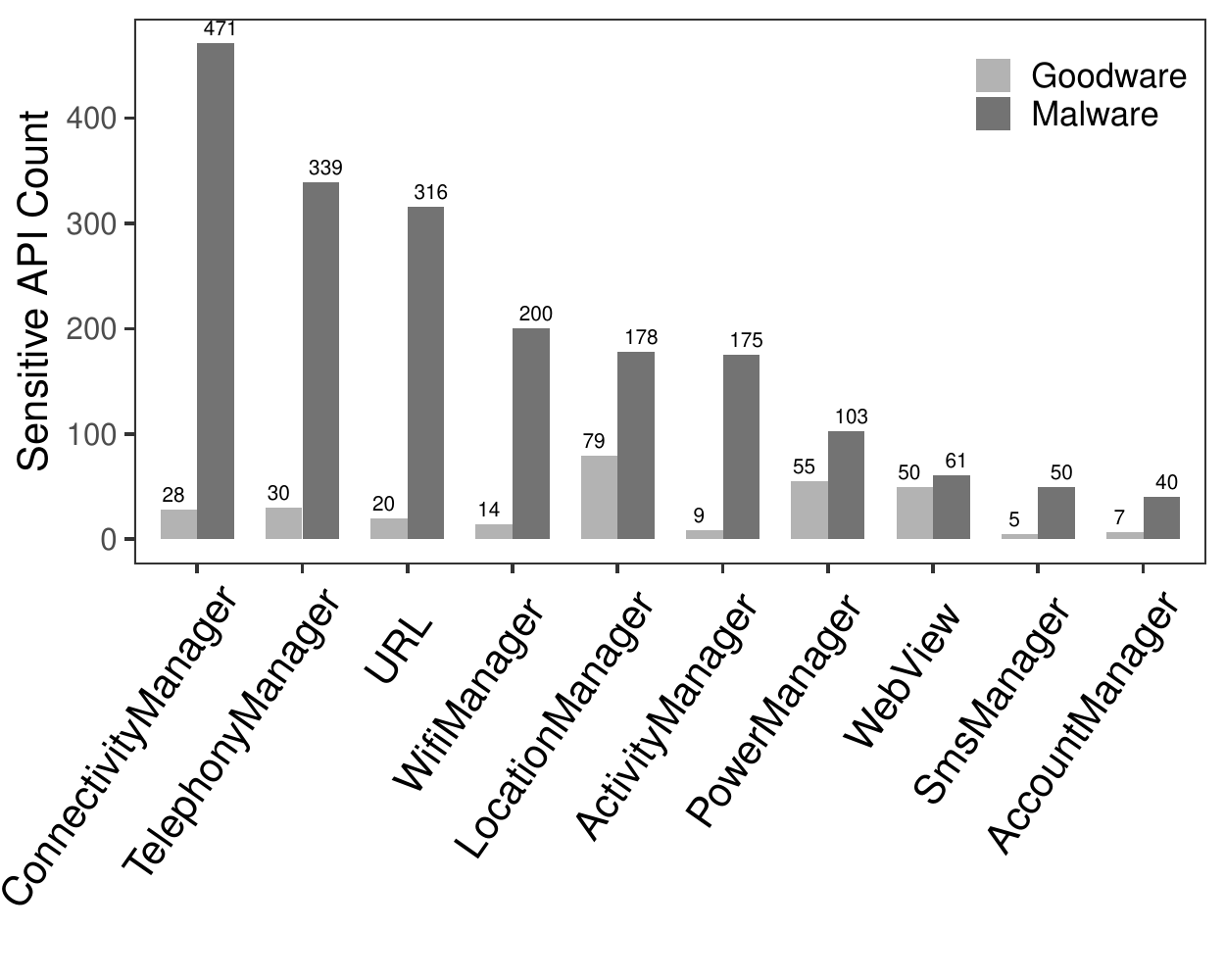}}
	\caption{Categories of Trigger Conditions in HSO and Sensitive APIs in Malware.}
    \label{fig:top10trigger_malware_goodware_sensitive_api}
\end{figure}

\subsection{Sensitive APIs involved in suspicious HSOs}

While the invocation of sensitive APIs does not necessarily mean it is malicious, sensitive APIs deliberately hidden in an HSB 
do raise its suspicion. 
\tool{} has identified 134 unique hidden sensitive APIs that appeared 3,195 times in our dataset. 
Figure \ref{fig:malware_goodware_sensitive_api} presents the top ten classes of the most frequently invoked sensitive APIs in \textit{malware set} and \textit{benign set}. The most involved APIs are network related, including the ones in \textit{URL}, \textit{ConnectivityManager}, and \textit{WifiManager} classes. The\textit{WebView} ((displays web pages))
and \textit{SmsManager} (manages SMS operations such as sending text messages) are also prevalently used in HSOs. Other commonly involved API classes include \textit{PowerManager} (controls the power state of the device such as keeping the screen stay on), \textit{LocationManager} (provides access to the system location services such as getting last known location), \textit{ActivityManager} (gives information about activities and services such as getting running tasks on the phone), \textit{TelephonyManager} (provides access to information of the telephony services such as phone number), and \textit{AccountManager} (manages user's online accounts). The detailed most common APIs in HSOs can be found in Table \ref{tab:top10apis}.

Interestingly, while most of the API classes have significantly more instances in malware samples than benign apps, \textit{WebView} is an exception. We therefore took an in-depth look into benign apps with WebView APIs in their HSOs and observed that 34 out of 50 cases are free apps that display advertisement web pages for revenue.

\begin{table}[!h]
\centering
\caption{Details of The Top 10 Classes of Hidden Sensitive APIs in HSO} 
\label{tab:top10apis}
\resizebox{0.65\linewidth}{!}{
\begin{tabular}{| r | l | } 
\hline
Class &  Most Frequent Sensitive API Examples \\
\hline
\multirow{3}{*}{ConnectivityManager}     &  
 net.ConnectivityManager\#getActiveNetworkInfo \\
 &   net.ConnectivityManager\#getNetworkInfo  \\
 &   net.ConnectivityManager\#getAllNetworkInfo  \\
\hline
\multirow{3}{*}{TelephonyManager}     & 
 telephony.TelephonyManager\#getDeviceId  \\
 &  telephony.TelephonyManager\#getSubscriberId \\
  &  telephony.TelephonyManager\#getCellLocation \\
\hline
\multirow{3}{*}{URL}       & 
 net.URL\#openConnection    \\
 &  net.URL\#getContent     \\
  &  net.URL\#openStream     \\
\hline
\multirow{3}{*}{WifiManager}     &  
  net.wifi.WifiManager\#getScanResults\\
 &    net.wifi.WifiManager\#getConnectionInfo  \\
  &    net.wifi.WifiManager\#getWifiState  \\
 \hline
 \multirow{3}{*}{LocationManager}       & 
 location.LocationManager\#getLastKnownLocation    \\
 &  location.LocationManager\#requestLocationUpdates \\
 &  location.LocationManager\#getBestProvider \\
\hline
\multirow{3}{*}{ActivityManager}     & 
 app.ActivityManager\#getRunningTasks  \\
 & app.ActivityManager\#getRecentTasks  \\
  & app.ActivityManager\#moveTaskToFront  \\
\hline
\multirow{3}{*}{PowerManager}     & 
 os.PowerManager.WakeLock\#release \\
 &  os.PowerManager.WakeLock\#acquire() \\
  &  os.PowerManager.WakeLock\#acquire(long) \\
\hline
\multirow{3}{*}{WebView}     & 
 webkit.WebView\#setBackgroundColor  \\
 &  webkit.WebView\#addJavascriptInterface   \\
  &  webkit.WebView\#loadDataWithBaseURL   \\
\hline
\multirow{3}{*}{SmsManager}       & 
 telephony.SmsManager\#sendTextMessage    \\
 &  telephony.SmsManager\#sendMultipartTextMessage  \\
 &  telephony.SmsManager\#sendDataMessage  \\
\hline
\multirow{3}{*}{AccountManager}     & 
accounts.AccountManager\#getAccountsByType \\
& accounts.AccountManager\#getAccounts \\
& accounts.AccountManager\#getUserData \\
\hline

\end{tabular} }
\end{table}

\subsection{Trigger Condition to Hidden Sensitive API Pairs}

\begin{figure}[!t]
    \centering 
    \includegraphics[width=0.8\linewidth]{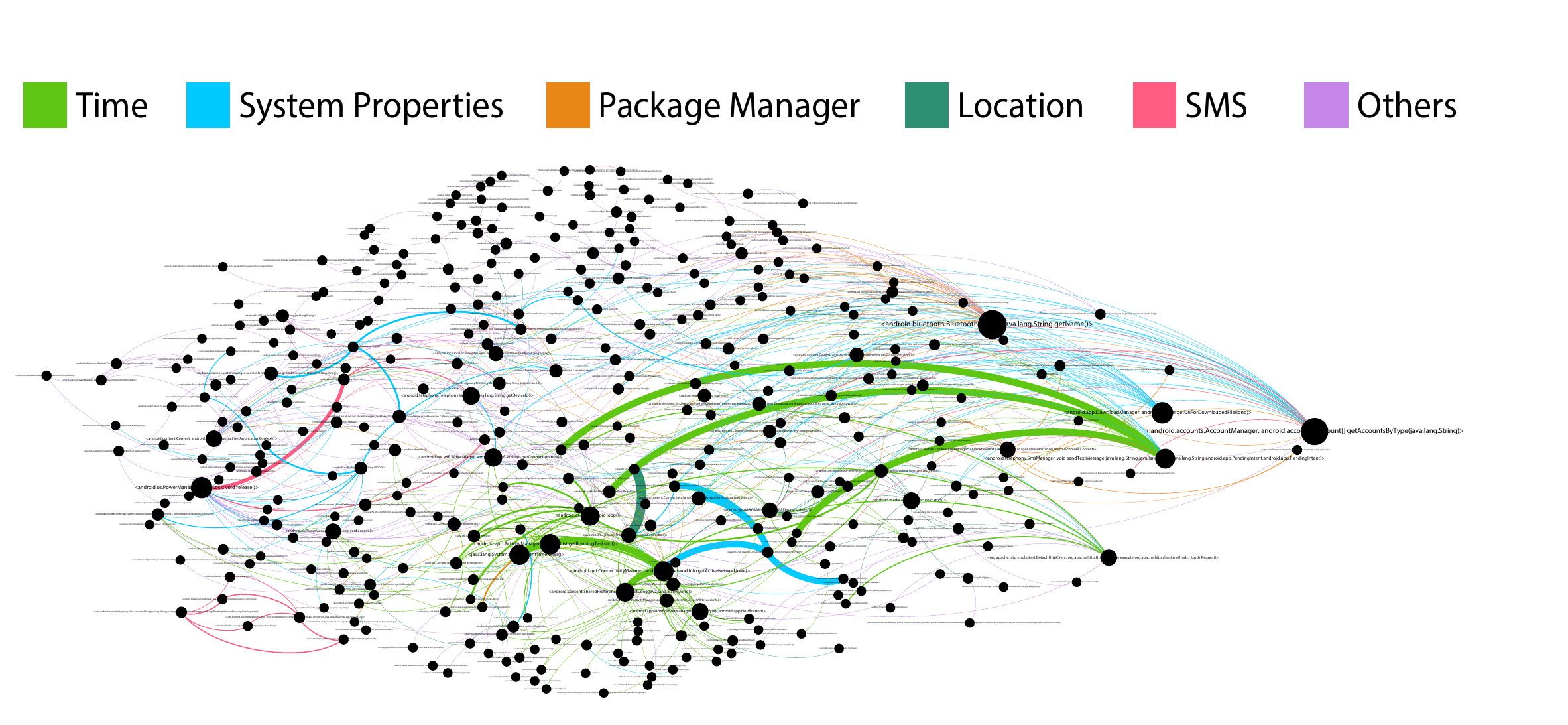}
    \caption{The HSO Trigger-Sensitive API Pairs Graph. }
    \label{fig:Trigger-Sensitive_Pair_graph}
\end{figure}

We now investigate the relationships between trigger conditions and the hidden sensitive APIs accessed in their corresponding HSOs so as to identify common patterns leveraged by attackers to achieve malicious purposes.
Figure~\ref{fig:Trigger-Sensitive_Pair_graph} graphically summarizes such relationships, i.e., trigger-to-hidden-sensitive-API pairs, where each node represents an API in either the trigger conditions or the hidden sensitive branches, while each edge denotes the connections between them. \tool{} has identified 404 nodes within which 346 are APIs in trigger conditions, 134 are APIs in hidden sensitive branches, and 15 APIs exist in both triggers and hidden sensitive branches. There are 2,847 edges found between them, which are illustrated in different colors according to their trigger conditions' categories. 

Table \ref{tab:top10pairs} further details the top 10 pairs found in HSOs with their categories and counts. The most frequent HSO patterns are to hide network-related activities behind retrieving the phone's location. For instance, the top one pattern that appeared 135 times in our dataset requests the SIM provider's country code. Based on the user's location, it then determines whether or not to open a web page, and what web pages (e.g., advertisement pages) to display to the user. 
Time-related HSO patterns are also widely found in the detected HSOs. They firstly compare the current system time with preset values. If the condition fulfills (e.g., the app is running for more than ten minutes), they try to initialize a network connection and send out the user's private information such as IMEI, phone number, etc. More than 250 instances in our dataset leverage this pattern to steal users' private information stealthily. Other frequent HSO patterns on the top list are involved in anti-emulator tricks include checking the phone's model name and checking if specific apps are installed (which could indicate if it is an emulator) before acquiring sensitive information.

\begin{table*}[!h]
\centering
\caption{Top 10 Trigger Condition to Hidden Sensitive API Pairs.}
\label{tab:top10pairs}
\resizebox{0.9\linewidth}{!}{
\begin{tabular}{ | l | l | l | c | } 
\hline
Category & Trigger Condition APIs & Hidden Sensitive APIs & Counts\\
\hline
Location &
TelephonyManager\#getSimCountryIso
& java.net.URL\#openConnection
& 135 \\
\hline
Location &
TelephonyManager\#getCellLocation 
& ConnectivityManager\#getActiveNetworkInfo
& 131 \\
\hline
Time &
java.lang.System\#currentTimeMillis 
& ConnectivityManager\#getNetworkInfo
& 80 \\
\hline
Time &
java.lang.System\#currentTimeMillis 
& ConnectivityManager\#getAllNetworkInfo
& 66 \\
\hline
Time &
java.lang.System\#currentTimeMillis 
& ConnectivityManager\#getActiveNetworkInfo
& 57 \\
\hline
System Properties &
android.os.Build\#MODEL
& TelephonyManager\#getSubscriberId
& 54 \\
\hline
System Properties &
android.os.Build\#MODEL
& Settings\$System\#putInt
& 54 \\
\hline
SMS &
android.os.Message\#obj
& PowerManager\$WakeLock\#release
& 52 \\
\hline
Package Manager & 
PackageManager\#getInstalledPackages
& ActivityManager\#getRunningTasks
& 51 \\
\hline 
Time &
java.lang.System\#currentTimeMillis 
& DefaultHttpClient\#execute
& 49 \\
\hline
\end{tabular} }
\end{table*}

\subsection{Suspicious HSOs in Third-party Code}
\label{subsec:library}

Finally, we further look into the identified HSOs to check if they are introduced by app developers or reused by third-party libraries.
For the sake of simplicity, we consider the code only located in the unique app package (also known as the app id) as developers newly implemented code while all the other code (i.e., in packages not connected with the app's domain name) as third-party code (e.g., third-party libraries).
Among the 2,201 HSOs, surprisingly, over half of them (i.e., 1,342) is contributed by third-party code (i.e., 1,173 HSOs in malware and 169 in benign apps), among which malware tends to be more favored to introduce HSOs through third-party code than benign apps.
This experimental evidence suggests that attackers have more incentives to achieve malicious behaviors through third-party code as it allows easy code reuse that makes it much easier to implement new malware.

\subsection{Comparison with state-of-the-art}
\label{subsec:comparison}

We now compare our approach with state-of-the-art works targeting the problem of detecting hidden sensitive operations. To the best of our knowledge, there are two closely related approaches: HSOMiner\cite{pan2017dark} and TriggerScope\cite{fratantonio2016triggerscope}. Unfortunately, the source code of HSOMiner is not publicly available, and it is infeasible to compare against it because they trained data on the authors' labelled dataset, which has also not been publicly released. We have contacted the authors about launching their approach to analyze Android apps. Unfortunately, we have not yet received any response from them. Similarly, the authors of TriggerScope have also not made it publicly available. As a result, we cannot compare with TriggerScope as well. Fortunately, Jordan Samhi has provided a re-implemented version\footnote{https://github.com/JordanSamhi/TSOpen} of TriggerScope based on the details given in its research paper. The re-implemented version is named as TSOpen (referring to as the open implementation of TriggerScope) and has already been leveraged by previous studies~\cite{samhi2021effectiveness}. In this work, we resort to comparing our approach with TriggerScope by actually comparing it with TSOpen. 

To set up the experiments for a fair comparison, we run TSOpen on the same 10,000 malware and 10,000 benign apps selected in evaluating \tool{} in section 4 (as indicated in the second and third columns in Table~\ref{tab:comparison_TS}). The experiments are executed under the same environment, i.e., the same server and the same timeout threshold (i.e., 20 minutes). 

The experimental results are summarized in Table~\ref{tab:comparison_TS}.
Overall, the number of HSOs found by \tool{} in goodware and malware (i.e., 441 and 1,790, respectively) is larger than those found by the TSOpen (i.e., 110 and 237, respectively). 
Recall that when evaluating the performance of \tool{} at the beginning of Section~\ref{sec:evaluation}, we have manually validated the 2,231 suspicious HSOs yielded by \tool{}, for which 1,938 are confirmed to be true positives, giving a precision of 86.8\%.
In this work, we further conduct the same manual validation for the results of TSOpen. 
Our manual validation confirms that TSOpen has at least correctly detected 90.2\% of logic bombs.
This result is expected as TSOpen only detects three types of HSOs (i.e., time, location, and SMS) while \tool{} aims at detecting a broader scope of HSOs.
To enable a fair comparison\footnote{We consider the original outputs of \tool{} and TSOpen for comparison since only a small number of their results could be false positive.}, in this work, we will only consider \tool{}'s results falling in these three categories.

As highlighted in Table~\ref{tab:comparison_TS} (cf. Columns 6-8), \tool{} detects more HSOs in all of the three categories. Among the detected HSOs, we find that 265 HSOs (186, 48, and 31 in time, location, and SMS, respectively) were detected by both tools (as summarized in the fourth row in Table~\ref{tab:comparison_TS}). Besides that, there are 802 HSOs (360, 357 and 85 in time, location, and SMS, respectively) exclusively detected by \tool{}, while still 82 HSOs (38, 1, and 43 in time, location, and SMS, respectively) identified by TSOpen are not flagged by \tool{}.

On a further investigation, we found the reason why \tool{} failed in detecting the 82 HSOs is that \tool{}'s definition of potentially-sensitive APIs is different from the definition in TSOpen. 
In this work, we consider all the APIs that are protected by Android permissions as potentially sensitive, while TSOpen takes a different approach to pre-select such a set of sensitive APIs\footnote{The sensitive APIs are a part of internal implementation of TSOpen, which is not configurable.}.
Their set of sensitive APIs includes both permission-protected and permission-free APIs.
For example, TSOpen treats the following two APIs, <android.content.BroadcastReceiver: void abortBroadcast()> and <android.os.Handler: boolean sendEmptyMessage(int)>, as sensitive APIs. However, \tool{} does not consider them as sensitive because they are not protected by permissions. 
Furthermore, considering that TriggerScope was published in 2016 and the Android API rapidly evolves, it is understandable that certain APIs (especially the latest ones) are not included, resulting in possibly less suspicious HSOs. 
Moreover, as claimed in their paper, TriggerScope only focused on characterizing logic bombs on some given behaviors, while \tool{} treated each sensitive API in state-of-the-art Android API-permission mappings~\cite{au2012pscout,backes2016demystifying,aafer2018precise,chaoran2022cross} as a target API, leading to better performance in terms of both quantity and variety in detected HSOs, compared with TriggerScope.

\begin{table}[t!]
\centering
\caption{The comparison results between \tool{} and TSOpen.} 
\label{tab:comparison_TS}
\resizebox{\linewidth}{!}{
\begin{tabular}{r | c c c c | c c c } 
\hline
Tool & \# Analyzed  & \# Analyzed & \# HSOs in & \# HSOs in & \# Time & \# Location & \# SMS  \\
Name & Goodware   & Malware & Goodware &  Malware & HSOs   & HSOs   & HSOs  \\
\hline
\tool{} & 10,000   & 10,000 & 441(in 322 apps) & 1,790(in 982 apps) & 546 & 405  & 116 \\
TSOpen & 10,000  & 10,000  & 110(in 51 apps)  & 237(in 123 apps)  & 229 & 49 & 69 \\
\hline
Common & 10,000  & 10,000  & 71  &  194 & 186 &  48 & 31 \\
\hline
\end{tabular} }
\end{table}

\subsection{Impact of Code Obfuscation}
\label{subsec:obfuscation}

As experimentally revealed by Zeng~{\cite{zeng2018resilient}} and Moser et al.~{\cite{moser2007limits}}, trigger conditions of HSOs could be obfuscated in order to evade the detection of advanced semantics-based malware analyzers.
Therefore, we are interested in checking to what extent our approach is impacted by obfuscation, especially when applied to pinpoint HSOs in real-world Android apps. 
Since there is no existing dataset that is suitable for our experiment, we resort to preparing such a dataset from scratch, i.e., to form a set of obfuscated app pairs for which each pair contains a non-obfuscated app and its obfuscated counterpart.
We start by randomly selecting 1,000 malware from our dataset and then apply Obfuscapk\cite{aonzo2020obfuscapk} on them to generate their obfuscated counterparts.
Obfuscapk is a modular Python tool designed to directly obfuscate closed-source Android apps.
Obfuscapk supports six types of obfuscation operations, which could be configured to achieve different granularities when obfuscating Android apps.

The six types of operations are summarized as follows.

\begin{enumerate}
\item Nop: Insert junk code. Nop, short for no-operation, is a dedicated instruction that does nothing. This technique just inserts random nop instructions within every method implementation.

\item Rename: operations that change the names of the used identifiers (classes, fields, methods).

\item Reorder: This technique consists of changing the order of basic blocks in the code. When a branch instruction is found, the condition is inverted (e.g., branch if lower than, becomes branch if greater or equal than) and the target basic blocks are reordered accordingly. Furthermore, it also randomly rearranges the code abusing goto instructions.

\item Reflection: This technique analyzes the existing code looking for method invocations of the app, ignoring the calls to the Android framework (see AdvancedReflection). If it finds an instruction with a suitable method invocation (i.e., no constructor methods, public visibility, enough free registers etc.) such invocation is redirected to a custom method that will invoke the original method using the Reflection APIs.

\item Advanced Reflection: Uses reflection to invoke dangerous APIs of the Android Framework. To find out if a method belongs to the Android Framework, Obfuscapk refers to the mapping discovered by Backes et al.~\cite{backes2016demystifying}

\item Encryption: packaging encrypted code/resources and decrypting them during the app execution. When Obfuscapk starts, it automatically generates a random secret key (32 characters long, using ASCII letters and digits) that will be used for encryption.
\end{enumerate}

In this work, we are interested in checking the impact of all of these six types of operations on our approach. Hence, for each of the selected apps and each obfuscation type, we launch Obfuscapk to generate an obfuscated app.
For the 1,000 selected apps, we expect to generate 6,000 obfuscated apps and eventually form 6,000 obfuscated app pairs.
We then launch \tool{} to analyze those apps and compare the number of detected HSOs obtained on apps with and without obfuscation.
Table~\ref{tab:comparison_obfuscation_malawre} summarizes our experimental results.

Expectedly, except for reflection, our approach is resilient to all the other four obfuscation types.
Our deep analysis reveals that the reason why \tool{} is unaffected by Nop obfuscator is that Nop obfuscator will only insert junk code, which is a dedicated instruction that does nothing. In terms of Rename and Reorder obfuscator, their code transformations will retain the functionality as the original APK thus will not impact our approach. Also, the reason why the Encryption obfuscator has no effect on \tool{} is that it will only encrypt constant strings in code, which will not impact the data flow analysis of our approach. In terms of reflection obfuscator and advanced Reflection obfuscator, both trigger conditions and sensitive API invocations can be redirected to 
other code entities by reflection calls, while those entities cannot be always resolved statically since the reflection call targets may not be statically resolved, which would lead to false negatives of \tool{}. 
The remaining two types that have an impact on our approach are all related to reflection, which performs complicated code changes that will likely break the data flow processes. 
Nevertheless, even for reflection, our approach can still detect around one-third of HSOs.

To better mitigate the impact of reflection-based obfuscation on our approach, we further propose to strengthen the capability of \tool{} by integrating the state-of-the-art reflection analysis tool DroidRA to handle reflection usages~\cite{sun2021taming}. After statically locating the reflective calls, DroidRA can transform a reflection-included Android app to a reflection-free version, where the located reflective calls will be represented by standard java calls. The newly generated reflection-free app would allow \tool{} to yield reflection-aware analysis results. Specifically, considering the 345 apps and 821 apps that are obfuscated by reflection calls and advanced reflection calls, respectively, we first apply DroidRA to convert them into 1,166 reflection-free apps. After that, we execute \tool{} to perform HSO analysis on these new apps and compare the number of detected HSOs obtained based on the original apps. As a result, \tool{} is able to detect all 13 reflection-relevant HSOs which are obfuscated with reflection obfuscation, while detecting 124 (with a success rate of 85.5\%) reflection-relevant HSOs  that are obfuscated with advanced reflection obfuscation. The reason why \tool{} fails on detecting a small portion of reflection-relevant HSOs is that DroidRA may not resolve all the advanced reflective calls. For example, DroidRA relies on COAL~\cite{octeau2015composite} solver to infer reflective calls, which might introduce false negatives, leading to reflection calls unresolved and thus can not be successfully detected by \tool{}. Nevertheless, our experimental result shows the capability of \tool{} in achieving most of the reflection-aware hidden sensitive operation detections.

\begin{table}[t!]
\centering
\caption{The comparison results of \tool{} before and after obfuscation techniques in malware.} 
\label{tab:comparison_obfuscation_malawre}
\resizebox{\linewidth}{!}{
\begin{tabular}{r | c c c c c c c c } 
\hline
\textbf{Obfuscator} & \textbf{Nop}  & \textbf{Rename} & \textbf{Reorder} & \textbf{Reflection} & \textbf{Advanced Reflection}  & \textbf{Encryption}\\
\hline
\# HSOs Before Obfuscation   & 144(in 821 apps) & 18(in 332 apps) & 146(in 792 apps)  & 13(in 345 apps)  & 145(in 821 apps)  & 141(in 811 apps)\\
\# HSOs After Obfuscation  & 144(in 821 apps)  & 18(in 332 apps)  & 146(in 792 apps) &  4(in 345 apps) &  64(in 821 apps)  & 141(in 811 apps)\\
\hline
Common & 144  & 18 & 146  &  4 & 64 &  141  \\
\hline
\end{tabular} }
\end{table}

\section{Implication: Detection of Hidden Sensitive Data Flows}
\label{sec:implication}
After being able to automatically detect suspicious HSOs, we now go one step further to investigate how such HSOs can bring security harms to users. There might be different security implications, in this work, we only focus on sensitive data leaks, which is also part of our initial attempts towards demonstrating the usefulness of identifying suspicious HSOs. Specifically, we are interested in detecting hidden sensitive data flows (HSDFs), i.e., leaking sensitive data collected through HSOs.
To the best of our knowledge, hidden sensitive data flow has not yet been explored by our community.
Unfortunately, it has not even been clearly defined.
To this end, we first define HSDF following the previous rules leveraged to define HSOs (cf. Section~\ref{sec:motivation}).
Let $S$ denote a sensitive data flow (also known as a private data leak as mentioned in the FlowDroid work~\cite{arzt2014flowdroid}), we consider that a sensitive data flow  happens when a sensitive ``tainted'' information goes from a source (e.g. the API method \emph{getDeviceId}) to a given sink (e.g. the API method \emph{sendTextMessage}).

\textbf{Definition 3 [Hidden Sensitive Data Flow (HSDF)]:}
A sensitive data flow $S$ is an HSDF if the source of $S$ appears in the hidden sensitive branch of a HSO.

Although HSDFs have not yet been specifically exploited by the state-of-the-art, our community has proposed various approaches to detect general sensitive data-flows.
One of the most famous approaches is FlowDroid~\cite{arzt2014flowdroid}, a state-of-the-art static analyzer that performs taint analysis to pinpoint sensitive data leaks flowing from a pre-defined set of \emph{source} methods to \emph{sink} methods.
These \emph{source} and \emph{sink} methods can be easily customized.
In this work, we leverage FlowDroid to detect sensitive data flows related to HSOs.
If a sensitive data flow reported by FlowDroid has its source method invoked in an HSO, we regard it as an HSDF.

By applying FlowDroid\footnote{In this work, the latest \emph{development} branch of FlowDroid{\cite{FlowDroid_develop_branch}} is leveraged for the experiments. It should be roughly equivalent to the FlowDroid 2.8 release.} to 1,304 apps (982 malware and 322 goodware) involving suspicious HSOs, we find that 67 apps further involve HSDFs, accounting to in total 401 HSDFs.
While manually checking the experimental results of FlowDroid and HiSenDroid, we find that 16 sensitive APIs, which are frequently invoked within HSOs to collect system information, are not taken into account by the source set of FlowDroid by default. These APIs (listed in Table~{\ref{tab:Hidden_source_methods}}), after manual confirmation, should still be considered as source methods by FlowDroid as they are responsible for retrieving sensitive data that should not be exposed to other parties. Here, to clarify, when doing the experiment, we include both of the default source and sink methods of FlowDroid and the additional sensitive APIs involved in HSOs in the SourceAndSink.txt file of FlowDroid. During the manual process, we have not found any sensitive API (i.e., involving dangerous operations) that should be additionally considered as a \emph{sink} method by FlowDroid.
Hence, we add the 16 APIs to the source set of Flowdroid and keep its sink set unchanged (hereinafter referred to this version as FlowDroid + \tool{}) and relaunch it on the same set of apps.
This time, we are able to disclose 1,110 HSDFs from 1,304 apps.
This result shows that suspicious HSOs could be leveraged to leak users' sensitive information outside of their devices.
As an example shown in Listing~\ref{code:SMS_Message}, the sensitive data \emph{device id} and \emph{subscriber id}, which are unique to the device and hence can be leveraged to uniquely track the phone, are eventually sent outside the device through a text message.

\begin{figure}[!h]
    \centering
    \includegraphics[width=0.65\linewidth]{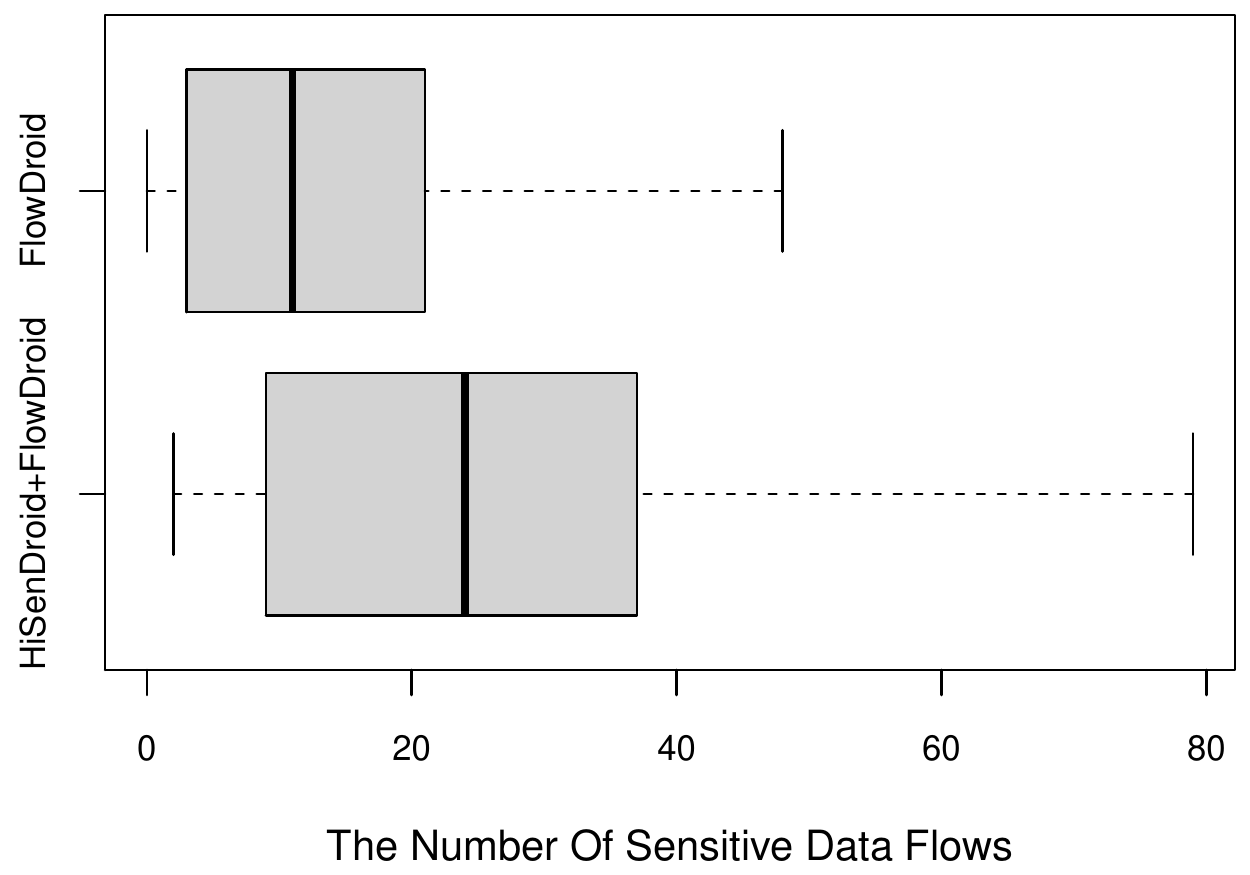}
	\caption{Results of Sensitive data Flows in Android Apps.}
    \label{fig:SDF_FlowDroid}
\end{figure}

Considering general sensitive data-flows (SDF), we compare FlowDroid with \tool{} on the same dataset. In general, among the 1,304 apps, HisenDroid+FlowDroid detect 31,215 SDF, which is significantly larger than that of the original FlowDroid (which is 16,946). This result, as expected\footnote{We remind the readers that, in this work, we did not improve FlowDroid by itself but only enlarged its \emph{source} set as some of the sensitive APIs, which are favored by HSOs, are overlooked by FlowDroid.}, does experimentally demonstrate the effectiveness of our approach towards revealing more data flows in Android Apps.
Our experimental results are illustrated in Figure~\ref{fig:SDF_FlowDroid}, which indicates the distribution of the number of sensitive data flows in each app yielded by HisenDroid+FlowDroid and \tool{}. This result shows that FlowDroid + \tool{}  has significantly improved the original results of FlowDroid, which shows the usefulness of our identified HSOs and suggests that there is a strong need to characterize hidden sensitive operations.

\begin{table}[!h]
\centering
\caption{The list of selected source methods that, by default, are not included by FlowDroid.}
\label{tab:Hidden_source_methods}
\resizebox{0.6\linewidth}{!}{
{\begin{tabular}{ l } 
\hline
API Signature \\
\hline
android.net.wifi.WifiManager\#getConnectionInfo()\\
android.app.ActivityManager\#getRunningTasks  \\
android.app.ActivityManager\#getRecentTasks   \\
android.accounts.AccountManager\#getUserData \\
android.net.ConnectivityManager\#getNetworkInfo  \\
android.provider.Settings\$System\#getUriFor  \\
android.telephony.TelephonyManager\#getNeighboringCellInfo \\
android.telephony.TelephonyManager\#getCellLocation  \\
android.accounts.AccountManager\#getAccountsByType  \\
android.net.wifi.WifiManager\#getScanResults  \\
android.net.wifi.WifiManager\#getConfiguredNetworks  \\
java.net.URL\#openConnection\\
android.net.ConnectivityManager\#getAllNetworkInfo \\
android.net.VpnService\#prepare  \\
android.hardware.Camera\#open   \\
android.net.ConnectivityManager\#getActiveNetworkInfo \\
\hline
\end{tabular} }}
\end{table}

\section{Limitations}
\label{sec:limitations}

The main limitation of our approach lies in the backward data-flow analysis, which applies only context-insensitive analysis and thereby may lead to imprecise results. Furthermore, at the moment, our approach is not aware of dynamically loaded code, reflectively accessed methods, and native code.
Subsequently, \tool{} may overlook certain app features and hence result in false-negative results.

Second, \tool{} data-flow analysis may be susceptible to obfuscation techniques. According to some former research works~\cite{glanz2020hidden, rasthofer2016harvesting, samhi2022jucify}, obfuscation (especially those involving complicated changes of the program code) may cause false negatives of the static analysis approach. Indeed, as demonstrated by Moser~\cite{moser2007limits}, obfuscation is actually a challenge for almost all static program analyzers. Just like all prior efforts on static analysis of HSOs ~\cite{fratantonio2016triggerscope}, ~\cite{pan2017dark}, we do not consider the apps whose branch conditions have been deeply obfuscated. 
Fortunately, the majority of obfuscations applied to Android apps only involve basic transformations (such as renaming~\cite{dong2018understanding}) that do not involve complicated code changes (e.g., structural or logic changes, or invoke sensitive code through reflections, etc.), which will not impact the analysis of our approach. This has also been confirmed by our exploratory study towards understanding the impact of obfuscation on our approach, as discussed in Section~\ref{subsec:obfuscation}. Considering reflection obfuscation, integrating DroidRA with \tool{} as a pipeline is demonstrated to be effective in eliminating the impact of reflection calls. Therefore, we believe that the technical capabilities and our results would not be significantly impacted by code obfuscation. Nevertheless, as part of our future work, we plan to integrate other approaches developed by our fellow researchers to mitigate these long-standing challenges, e.g., by applying DroidRA~\cite{sun2021taming,li2016droidra} to mitigate the impact of reflection-enhanced code obfuscations.

Although summarized from many sensitive operations, the definition of HSO rules may not be perfect. Indeed, on the one hand, the set of sensitive operations considered for summarization may not be representative, and the set of apps leveraged to obtain such sensitive operations may not be represented as well. On the other hand, the manual analysis leveraged to summarize the rules may contain errors since it is known that human efforts are prone to errors. Apart from that, the definition of HSO is based on empirical evidence that might not be perfect. There might be complicated cases that do not follow the definition but still manifest themselves as hidden sensitive behaviors in practice, leading to false negatives. This limitation can also apply to the conventional usage analyses since the list of conventional usages is manually summarized based on a given set of apps. The subsequent outputs (i.e., whitelist) may not be representative. Nonetheless, our follow-up study using a set of 20,000 new apps has shown that this impact is negligible. Furthermore, in this work, we have attempted to provide detailed insights to explain why HSOs are reported as such. This knowledge is expected to be useful for practitioners and researchers to characterize conventional usages and for security analysts to understand suspicious HSOs.

Moreover, since the original implementation of TriggerScope is not publicly available, we have resorted to an open re-implementation version of TriggerScope to compare our approach against it. This alternative decision may result in possible biases as the re-implementation may not really represent the original version. Unfortunately, the re-implemented version is the only source we can publicly locate to fulfill the comparison.  As of our future work, we plan to also evaluate the reliability of the re-implementation of TriggerScope so as to mitigate potential biases, if any.

Last but not the least, the performance of the hidden sensitive data flow analysis may be impacted by the collection of sensitive APIs (i.e., sources). On one hand, some sensitive APIs, especially the latest ones, might be overlooked by FlowDroid and hence cannot be considered for pinpointing potential leaks, leading to false-negative results. In this work, our experimental results have confirmed this. On the other hand, some historical sensitive APIs included in FlowDroid's source list might be deprecated and subsequently removed from a certain Android API version~\cite{li2018cid}. There is hence no need to include them when analyzing apps targeting higher API versions, as these APIs will not be used anymore, not even mentioning causing sensitive data leaks. To overcome these impacts, we believe there is a need to keep updating FlowDroid's list of sensitive APIs, in order to achieve a more effective and sound sensitive data flow analysis for Android apps. Furthermore, ideally, FlowDroid should also not be expected to include APIs that are released after itself.

\section{RELATED WORK}
\label{sec:related_work}
Hidden sensitive operations have long existed in Android malware as evasive technologies have widely been used by attackers to hide their malicious behaviors.
Our research community has hence proposed various approaches to tackle these issues. We now discuss some of the representative works from two angles, including the evasive techniques that have been proposed to hide malicious code from being identified, and the detection methods proposed to pinpoint such evasive techniques.

\textbf{Evasive Techniques.} There has been a number of research works on hiding malicious behavior from detection, most of which focus on evading the dynamic test platforms such as virtual machines and emulators. Early works target the Windows platform \cite{chen2008towards}, while recently the trend has been moved to Android \cite{petsas2014rage,vidas2014evading,jing2014morpheus,diao2016evading,costamagna2018identifying,liu2021deep,liu2022explainable}. These evasive techniques detect the presence of a simulated environment by either looking into the system properties of the testing platform (e.g., system fingerprints, hardware capabilities, etc.) \cite{petsas2014rage,vidas2014evading,costamagna2018identifying}, or leveraging a reverse Turing test that examines if the app interacts with a human user \cite{diao2016evading}. For instance, Diao et al. \cite{diao2016evading} observed that programmed interaction has specific patterns of input and interaction frequency, which is different from real users. Overall, the evasive techniques usually hide malicious activities in an \emph{if-then-else statement}. The hidden malicious behavior will only be set off when certain conditions are fulfilled (e.g., not in an emulator); otherwise dummy benign operations are triggered. The prevalence of such evasive techniques motivated us to investigate the HSOs in Android apps and propose \tool{} to detect them.

\textbf{Detection of Evasive Techniques.} The pervasive evasive techniques (e.g., anti-emulator techniques) have motivated the research community to take countermeasures. Great effort has been spent on detecting known types of hidden behaviors that hampers the dynamic analysis process. These works include detecting anti-emulator techniques \cite{balzarotti2010efficient, lindorfer2011detecting, kirat2014barecloud,kirat2015malgene} and generic logic-bombs \cite{crandall2006temporal,brumley2008automatically,zheng2012smartdroid,fratantonio2016triggerscope,papp2019towards}. The approaches of detecting anti-emulator techniques compare the behavioral deviation of the tested apps on the various environments when feeding them the same input. The fundamental idea is that if the app behaves differently in different environments, it is likely trying to evade one or more analysis platforms (usually referred to as bare-metal analysis in the literature) \cite{balzarotti2010efficient, lindorfer2011detecting, kirat2014barecloud,kirat2015malgene}. While these early works investigate a critical category of hidden operations (i.e., anti-emulator), the proposed methods lack generalization that cannot be applied to detect other types of hidden operations emerging recently.

Besides the detection of anti-emulator techniques, several works are focusing on uncovering other trigger-based behaviors. These approaches leverage symbolic execution or static code analysis and instrumentation to expose the hidden branches in an \emph{if-then-else statement} \cite{crandall2006temporal,brumley2008automatically,zheng2012smartdroid,fratantonio2016triggerscope,papp2019towards}. As examples, Zheng et al. \cite{zheng2012smartdroid} proposed to leverage a static analysis approach to retrieve all UI related events, and use dynamic testing to trigger them and log the invocation of sensitive APIs. Unlike \tool{} that leverages static analysis, the dynamic analysis based approach introduces significant system- and time-overhead. The coverage of the dynamic analysis is also in question.
Fratantonio et al. \cite{fratantonio2016triggerscope} proposed TriggerScope to detect hidden triggered behaviors based on the observation that certain triggers (i.e., time, location, and SMS related triggers) always involve the comparison of specific types of input (i.e., system time, system location, and received SMS). Symbolic execution is then leveraged to detect such narrow conditions. While TriggerScope is effective in detecting the above-mentioned three types of logic bombs, it cannot be generalized to detect hidden operations triggered by other types of conditions, such as system property, which has been found pervasive in Android apps.

Similar to \tool{}, another line of work attempts to detect unknown types of trigger-based behaviors~\cite{pan2017dark},~\cite{wang2017droid}.
A prominent example is HSOMiner~\cite{pan2017dark}, which extracts static characteristics of hidden behaviors as features and trains a machine learning model to identify the code blocks that observe similar patterns. The major differences between our work and HSOMiner are twofold. First, HSOMiner requires a large number of manually labelled training samples, which involves extensive human experts' effort. Its performance also heavily relies on the manually labelled training data, which is prone to errors. Our method, on the other hand, is an automatic process without human intervention. Second, HSOMiner, as a machine learning based approach, lacks explanations of the decisions. In contrast, our static code analysis based approach outputs the full call traces of detected HSOs, and provides more detailed information for further analysis.

\section{CONCLUSION}
\label{sec:conclusion}

In this work, we present to the community a prototype tool called \tool{}, which performs a static code analysis to uncover hidden sensitive operations that will only be triggered under special circumstances such as at a specific location or in a certain time period.
Additionally, \tool{} goes one step deeper to provide details aiming at helping security analysts understand why a given hidden sensitive operation is flagged as such.
Experimental results over 20,000 apps, including both malicious and benign apps, show that hidden sensitive operations are indeed quite frequently presented in Android apps and \tool{} is effective in automatically discovering them.
Moreover, with the help of FlowDroid, a state-of-the-art static taint analyzer, we further experimentally find that hidden sensitive operations could eventually lead to privacy leaks.

\section{Acknowledgments}

The authors would like to thank the anonymous TOSEM reviewers who have provided insightful and constructive comments, which have been extremely useful for helping in improving this manuscript.
This work was partly supported by the Australian Research Council (ARC) under a Laureate Fellowship project FL190100035, a Discovery Early Career Researcher Award (DECRA) project DE200100016, and a Discovery project DP200100020, 
by the Luxembourg National Research Fund (FNR) (under project CHARACTERIZE C17/IS/11693861), by the SPARTA project, which has received funding from the European Union's Horizon 2020 research and innovation program under grant agreement No 830892. 

\balance
\bibliographystyle{acm}
\bibliography{ref}

\end{document}